\documentclass[final,5p,times,twocolumn]{elsarticle}

\usepackage{amssymb}
\usepackage{amsmath}
\usepackage{amsthm}
\usepackage{amsfonts}
\usepackage{color}
\usepackage{epstopdf}

\usepackage{subcaption}
\usepackage{graphicx}
\usepackage{bbm}
\usepackage{dsfont}
\usepackage{braket}

\usepackage{stfloats}

\usepackage{mathtools,etoolbox}
\DeclarePairedDelimiterX{\abs}[1]{\lvert}{\rvert}{\ifblank{#1}{{}\cdot{}}{#1}}

\newcommand{\tcb}{\textcolor{black}}
\newcommand{\tcr}{\textcolor{black}}

\DeclareMathOperator{\sign}{\operatorname{sign}}

\usepackage[english]{babel}

\journal{Journal of Neuroscience Methods}

\begin{document}

\begin{frontmatter}

\author[UniFi]{Arturo Mariani}
\author[UniFi]{Federico Senocrate}
\author[MU]{Jason Mikiel-Hunter}     % jason.mikiel-hunter@mq.edu.au
\author[MU]{David McAlpine}  		    % david.mcalpine@mq.edu.au
\author[GS-LMU]{Barbara Beiderbeck}  % beiderbeck@biologie.uni-muenchen.de
\author[LMU]{Michael Pecka}  			% pecka@bio.lmu.de
\author[ENSEA]{Kevin Lin}
\author[ISC,INFN]{Thomas Kreuz}
\ead{thomas.kreuz@cnr.it}

%% \ead[url]{home page}

% ##### Please check your affiliations #####
\address[UniFi]{Department of Physics and Astronomy, University of Florence, Sesto Fiorentino, Italy}
\address[MU]{Department of Linguistics, Macquarie University, Sydney, Australia}
\address[LMU]{Division of Neurobiology, Faculty of Biology, Ludwig-Maximilians-Universität, Munich, Germany}
\address[GS-LMU]{Graduate School of Systemic Neurosciences, Ludwig-Maximilians-Universität, Munich, Germany}
\address[ENSEA]{École Nationale Supérieure de l'Électronique et de ses Applications, Cergy, France}
\address[ISC]{Institute for Complex Systems (ISC), National Research Council (CNR), Sesto Fiorentino, Italy}
\address[INFN]{National Institute of Nuclear Physics (INFN), Florence Section	, Sesto Fiorentino, Italy}

\title{Latency correction in sparse neuronal spike trains with overlapping global events}

\begin{abstract}

% XXXXX Must be less than 250 words XXXXX
\textit{Background:} 
In Kreuz et al., J Neurosci Methods 381, 109703 (2022) two methods were proposed that perform latency correction, i.e., optimize the spike time alignment of sparse neuronal spike trains with well-defined global spiking events. The first one based on direct shifts is fast but uses only partial latency information, while the other one makes use of the full information but relies on the computationally costly simulated annealing. Both methods reach their limits and can become unreliable when successive global events are not sufficiently separated or even overlap. 
% directly extracted from delays between neighboring spike trains 

\noindent \textit{New Method:} 
Here we propose an iterative scheme that combines the advantages of the two original methods by using in each step as much of the latency information as possible and by employing a very fast extrapolation direct shift method instead of the much slower simulated annealing.

\noindent \textit{Results:} 
We illustrate the effectiveness and the improved performance, measured in terms of the relative shift error, of the new iterative scheme not only on simulated data with known ground truths but also on single-unit recordings from two medial superior olive neurons of a gerbil.

\noindent \textit{Comparison with Existing Method(s):} 
The iterative scheme outperforms the existing approaches on both the simulated and the experimental data. Due to its low computational demands, and in contrast to simulated annealing, it can also be applied to very large datasets.
% XXXXX In case of no overlap the method falls back on the original approach XXXXX True? Should be because stop diagonal N-1 works and no extrapolation is needed XXXXX

\noindent \textit{Conclusions:} 
The new method generalizes and improves on the original method both in terms of accuracy and speed. Importantly, it is the only method that allows to disentangle global events with overlap. 

\end{abstract}

%\begin{keyword}
%	spike train analysis \sep latency correction \sep SPIKE-synchronization \sep SPIKE-Order \sep Synfire Indicator
%\end{keyword}

\end{frontmatter}
%% \linenumbers

% #############################################################################
% ######################## Section: Intro #####################################
% #############################################################################
%
\section{Introduction} \label{s:Intro}
%
%Overall structure of introduction:
%
% Discrete data, Rasterplot (examples)
% 
% Existing approaches for quantifying synchronization and directionality
% based on spike matching (SPIKE-synchronization, SPIKE-Order)
% 
% Our latest approach: Latency Correction
%
% Particularly useful for data with global events (examples)
% 
% Underlying criterion: Adaptive coincidence detection, Condition based on closeness in time, usually very reasonable but there is one problem: 
% 
% What do you do when these global events overlap?
%
%- Our approaches
%
%- Short overview of the study (The remainder of this paper ...)
%
% XXXXX \citep{wongmassang2021weakly} XXXXX

% Discrete data (spike trains) and Rasterplots; Two possible scenarios
The recording of neurophysiological data typically involves the high frequency sampling of one or more continuous signals. Often in a second step these quasi-continuous time series are transformed into discrete point processes. The most prominent example is neuronal spike detection: Under the assumption that neither the shape of the spike nor the background activity carry relevant information, single-unit responses are reduced to a spike train where the only information maintained is the timing of the individual spikes \citep{hennig2019scaling}.

% Two scenarios (here only in passing, was much longer in last paper)
A set of such spike trains can best be represented in a \textit{rasterplot} where the time stamps of the discrete events are plotted versus time. Here rows can represent either a population of different neurons recorded simultaneously or successive recordings of the same neuron that are aligned relative to some common trigger event (e.g., the onset times of repeated presentations of a stimulus).

\begin{figure*}[!ht]
	\begin{center}
	\includegraphics[width=\linewidth]{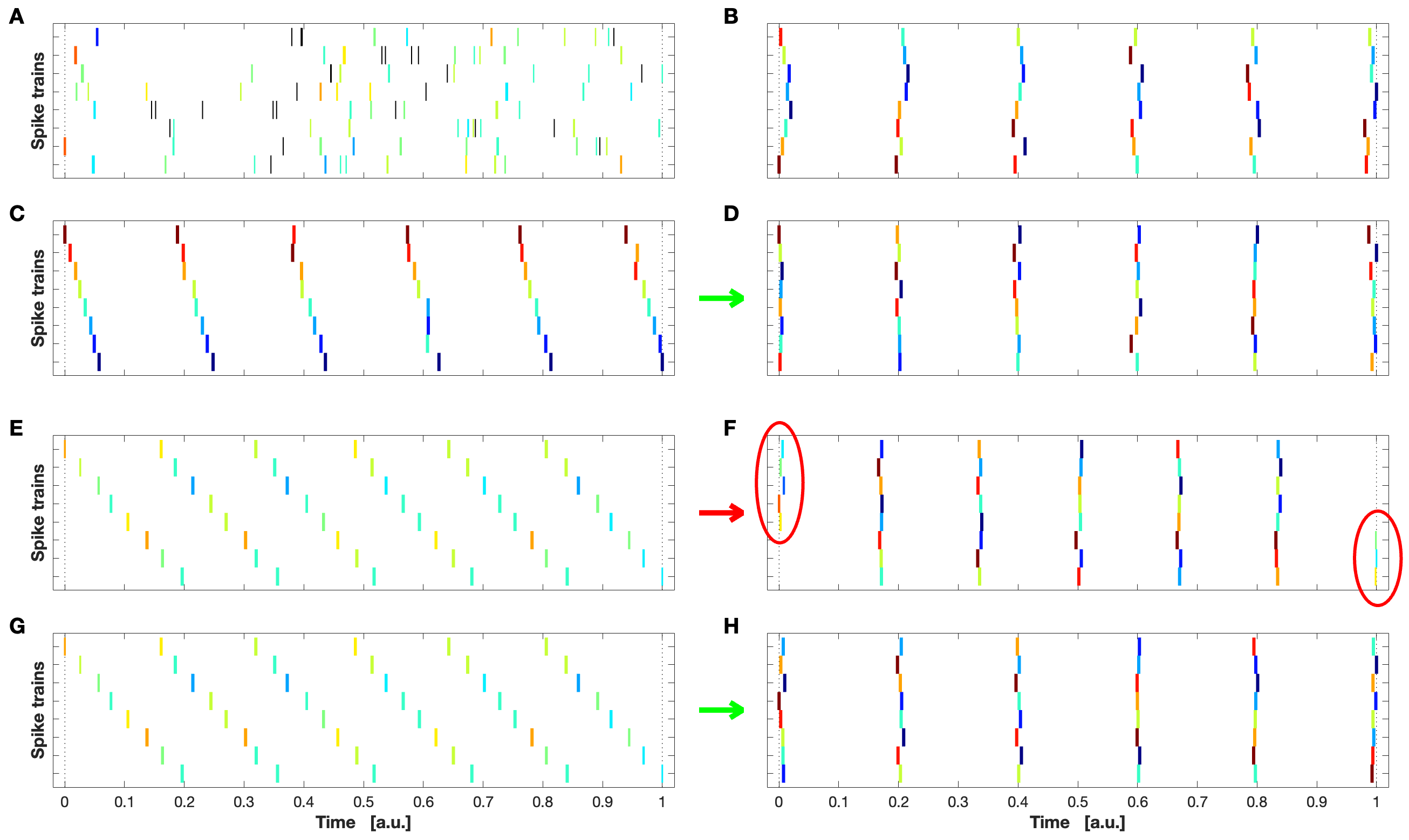}
	\caption{A. Spike train rasterplot with low SPIKE-synchronization (and accordingly low Synfire Indicator). B. Maximum SPIKE-synchronization but low Synfire Indicator. C. Spike trains with maximum SPIKE-synchronization and high Synfire Indicator. This corresponds to an almost perfect synfire chain without any overlap between successive events. D.  Latency correction of C yields almost perfectly identical spike trains (just a bit of jitter). E. Synfire chain with overlap. F. Standard latency correction of E results in spurious mismatches. G. Synfire chain with overlap (exactly the same as in E). H. The improved latency correction yields perfectly aligned spike trains without any mismatches. Spike colors code the anti-symmetric SPIKE-Order $D$ (see Section \ref{ss:Methods_1-2_Spike-Sync-Order-Synfire-Indicator} and \ref{App3-SPIKE-Order-Synfire-Indicator}) on a scale from $1$ (red, first leader) to $-1$ (blue, last follower).}
	\label{Fig1:Overview-Rasterplots}
	\end{center}
\end{figure*}

% Global events
Rasterplots can consist of rather steady and continuous firing with a fairly flat spike time histogram (such as the example in Fig. \ref{Fig1:Overview-Rasterplots}A) but often they exhibit pronounced global events in which case \tcr{all or at least a very large majority of spike trains fire within a rather short time window and accordingly,} the spike time histogram shows very prominent peaks. Such events can appear either due to spatiotemporal propagation where activity starts somewhere and then spreads over the whole recording region or within the context of neuronal coding when repeated presentations of a stimulus result in similar but typically not identical responses (e.g., due to variations in onset latency).

% SPIKE-synchronization, Applications (Relevance)
In recent years a number of new measures have been proposed that are particularly suited to quantify the synchronization and the directionality contained in these global events. The symmetric measure SPIKE-synchronization \citep{Kreuz15} quantifies the degree of synchrony from the relative number of quasi-simultaneous appearances of spikes. An example of a rasterplot with well matched spikes exhibiting maximum SPIKE-synchronization is shown in Fig. \ref{Fig1:Overview-Rasterplots}B. SPIKE-synchronization has, among others, been employed to investigate neuronal coding in the primary auditory cortex in echolocating bats \citep{macias2020temporal}, to evaluate the role of hub neurons in modulating cortical dynamics \citep{gal2021role}, and to quantify the effect of electrical stimulation in neural fibers \citep{hussain2024highly}. 

% SPIKE-Order, Synfire Indicator, Applications (Relevance)
SPIKE-synchronization is complemented by SPIKE-Order \citep{Kreuz17}, an anti-symmetric directionality approach that allows to sort spike trains from leader to follower and to quantify the consistency in the underlying spatio-temporal propagation patterns via the Synfire Indicator \citep{Kreuz17}. While the example from Fig. \ref{Fig1:Overview-Rasterplots}B is not consistent in its spike order, the order within the rasterplot of Fig. \ref{Fig1:Overview-Rasterplots}C is almost perfectly consistent and thus yields a Synfire Indicator close to its maximum value. SPIKE-Order has been applied to datasets such as invasive EEG recordings in children with refractory epilepsy \citep{tomlinson2019reproducibility}, responses of the inferotemporal area of a monkey upon visual stimulation \citep{gainutdinov2021method} and cortical activation patterns recorded via wide-field calcium imaging from mice before and after stroke \citep{Cecchini21}.
%The most recent application concerns the recognition of sounds by ensembles of proteinoids \citep{mougkogiannis2024recognition}.

% 
% ###########################################################################
% ################## Figure 1: Rasterplot Examples ##########################
% ###########################################################################
%

% Latency Correction
Most recently, both of these measures proved very useful as criteria for new algorithms designed to eliminate latency in sparse neuronal datasets \citep{Kreuz22}. The underlying idea is that for the estimation of synchrony latency is actually a hindrance and it is by optimally aligning spike trains \tcr{that their synchrony can be judged independent of any delays that might be present} \citep{Kreuz22}. For example, latency correction shows that the spike trains forming the synfire chain of Fig. \ref{Fig1:Overview-Rasterplots}C, when correctly aligned, are actually almost identical (Fig. \ref{Fig1:Overview-Rasterplots}D). The algorithm uses simulated annealing to minimize the distances between spikes and thereby eliminates any systematic delays but maintains all other kinds of noisy disturbances in the data. 

% Coincidence criterion: Closeness in time
As an essential prerequisite for all of these approaches we need a criterion to determine which spikes of a given rasterplot should be compared against (and then aligned with) each other. This need is met by spike matching via the adaptive coincidence criterion originally introduced for the bivariate measure event synchronization \citep{QuianQuiroga02b}. According to this criterion two spikes are considered coincident when they lie in each other's coincidence window. The length of each window is defined relative to the local firing rate of the respective spike train: A high firing rate (small interspike intervals) leads to a small window, whereas sparse firing (long interspike intervals) increases the window. 

% Problem: Overlap
So the relevant criterion is ‘‘closeness in time" but closeness is not defined by a fixed interval but rather in a more flexible manner adapted to the local properties of the data. For a dataset with global events this means that all the spikes within such an event are coincident with each other if and only if the interval between successive events is at least twice as large as the propagation time within an event \tcr{(i.e., if the data are sufficiently sparse)}. Under normal circumstances this criterion of closeness in time is very reasonable but problems start to arise as soon as successive global events are not sufficiently separated or even overlap (e.g., when the following event starts before the previous one has ended, as in Fig. \ref{Fig1:Overview-Rasterplots}E). Such overlap results in ‘‘mismatches" where spikes from different global events are paired together, which then in turn leads to unintended consequences in the estimation of SPIKE-synchronization and the Synfire Indicator as well as spurious shifts in the latency correction (as shown in Fig. \ref{Fig1:Overview-Rasterplots}F).

% Aim and scope of the paper
\tcr{The main objective of this article} is to propose an algorithm that deals with overlapping events in such a way that the intended spike matching is achieved (e.g., all spikes from the same true global events are synchronous to each other and inter-event matchings are minimized). \tcr{This kind of analysis allows to judge the faithfulness of activity propagation between different neurons or between different brain areas \citep{Kumar10} or, in the context of neuronal coding, to estimate to what extent the response to repeated presentation of a stimulus is independent of variations in onset latency \citep{levakova2015review}.} The \tcr{fundamental} idea is to use an iterative scheme. We start with an initial spike matching based on adaptive coincidence detection that includes only pairs of spikes that are not affected by overlap. These matches are then used in a first latency correction that improves the alignment of the spike trains and thereby ideally eliminates any overlap that might have been present in the original spike trains. Subsequently, we use these newly aligned spikes in a second iteration of spike matching and latency correction. This procedure not only leads to a better latency correction (Fig. \ref{Fig1:Overview-Rasterplots}G, H) but also to improved estimates of SPIKE-synchronization (or any other measure of synchrony). In this study we will investigate the performance of this iterative scheme on both simulated data with known ground truths and on single-unit recordings from two medial superior olive (MSO) neurons of a gerbil. These data are recorded during encoding of interaural time differences (ITD), i.e., the differences in arrival times of noisy sounds at the two ears. They exhibit global events with overlap and are thus a perfect testing ground for our new algorithm.

% Structure of the article
The remainder of the article is organized as follows: In Section \ref{s:Data} we describe the gerbil data that we use to illustrate the effectiveness of the algorithm at work under real-life conditions. The Methods are divided into two Sections. First, in Section \ref{s:Methods1} we describe the work that has been done so far. This includes spike matching through adaptive coincidence detection (Section \ref{ss:Methods_1-1_Coincidence-Detection}), the SPIKE-Order framework (Section \ref{ss:Methods_1-2_Spike-Sync-Order-Synfire-Indicator}) and the recently proposed algorithms for latency detection in sparse data (Section \ref{ss:Methods_1-3_Latency-Correction}). In Section \ref{s:Methods2} we introduce the problem of event overlap (Section \ref{ss:Methods_2-1_Overlap-Theory}) and propose a new approach to tackle it (Section \ref{ss:Methods_2-2_New-Algorithms}) as well as a novel way to quantify our success in doing so (Section \ref{ss:Methods_2-3_Relative-Shift-Error}). In the Results (Section \ref{s:Results}) we apply this new approach to artificially generated datasets (Section \ref{ss:Results-Sim}) and to the neurophysiological gerbil recordings (Section \ref{ss:Results-Exp}). Conclusions are drawn in Section \ref{s:Conclusions}. Finally, in the Appendix we provide a description of the experimental setup and the data collection (\ref{App1-Experimental-Setup}), recapitulate SPIKE-Synchronization (\ref{App2-SPIKE-synchronization}) as well as SPIKE-Order and the Synfire Indicator (\ref{App3-SPIKE-Order-Synfire-Indicator}) in more detail and derive our new relative shift error metric in \ref{App4-Relative-shift-error}.

% #############################################################################
% ######################### Section: Data #####################################
% #############################################################################
%
\section{Data} \label{s:Data}

\tcb{In this article we will investigate the performance of the new latency correction algorithm on experimental datasets that were recorded to investigate the neural coding of interaural time differences (ITDs), i.e., how small differences in the arrival times of sounds between the two ears of a mammal enable the spatial localization of the noise source. Here we provide a short overview of the experimental setup and the data collection. For the more technical details on the recording process please refer to \ref{App1-Experimental-Setup}.}

\tcb{The analyzed data consist of \textit{in vivo}, single-unit recordings from two Medial Superior Olive (MSO) neurons in a juvenile Mongolian Gerbil (Meriones Unguiculatus) \cite{Beiderbeck22} to whom $10$ different frozen white noise tokens of $100$ms duration ($5$ms rise/fall times) were presented binaurally via earphones at intensities $30$dB above a neuron’s threshold. The noise tokens were played at $61$ ITDs between $-1.79$ms and $1.79$ms which were introduced by advancing or delaying a stimulus by half the required ITD value at either ear. Every ITD and noise token combination was repeated $3$ times in a pseudo-random fashion, leading to a total of $183$ trials in each of the $10$ rasterplots reconstructed for the two neurons ($20$ datasets in total).}

% As binaural coincidence detectors of low-frequency, phasic signals, the rasterplots of MSO neurons’ global spiking events demonstrate an ITD-dependent latency which in each ITD condition exhibits strong overlap with the inter-event intervals of the underlying monaural inputs.
%
% Since different interaural time differences lead to variations in latency, the resulting rasterplots contain global spiking events with varying latencies and due to the vicinity of event-inducing triggers in the noise stimuli these global events often exhibit overlap. 
%
\tcb{Since MSO neurons act as binaural coincidence detectors of low-frequency, phasic signals, the global spiking events within their rasterplots demonstrate an ITD-dependent latency which in each ITD condition exhibits strong overlap with the inter-event intervals of the underlying monaural inputs. It is exactly this property (overlapping global events) which renders these datasets perfect test cases for our new latency correction algorithm. Moreover, the precise timing of the stimulus delivery allows us for ever spike train to determine the exact physiological delay caused by the ITDs. With this knowledge we can produce for each of the $20$ recordings a \emph{compensated} version of the rasterplot without any systematic delays which will then serve us well as benchmark (or ground truth) for the latency correction.} 

% #############################################################################
% ###################### Section: Methods #####################################
% #############################################################################
%
\section{Methods I - Previous work based on spike matching} \label{s:Methods1}

First, we introduce the matching of spikes by means of adaptive coincidence detection as the fundamental step underlying all measures and algorithms discussed in this paper (Section \ref{ss:Methods_1-1_Coincidence-Detection}). Subsequently, in Section \ref{ss:Methods_1-2_Spike-Sync-Order-Synfire-Indicator}, adaptive coincidence detection is used to quantify the overall level of synchronization using the measure SPIKE-synchronization and to analyze leader-follower relationships in a spike train set via SPIKE-Order and the Synfire Indicator. Adaptive coincidence detection also forms the basis of recently proposed algorithms to search for systematic delays in sparse neuronal spike trains and deal with them by means of latency correction (Section \ref{ss:Methods_1-3_Latency-Correction}).

\subsection{Spike matching via adaptive coincidence detection} \label{ss:Methods_1-1_Coincidence-Detection} 

Adaptive coincidence detection was first proposed for event synchronization in \cite{QuianQuiroga02b} and has by now been used in more than one hundred articles \footnote{For a constantly updated lists of these studies please refer to https://www.thomaskreuz.org/publications/isi-spike-articles.}.
% 
% #############################################################################
% ################## Figure 2: Coincidence Detection ##########################
% #############################################################################
%
\begin{figure}[!ht]
	\begin{center}
	\includegraphics[width=\linewidth]{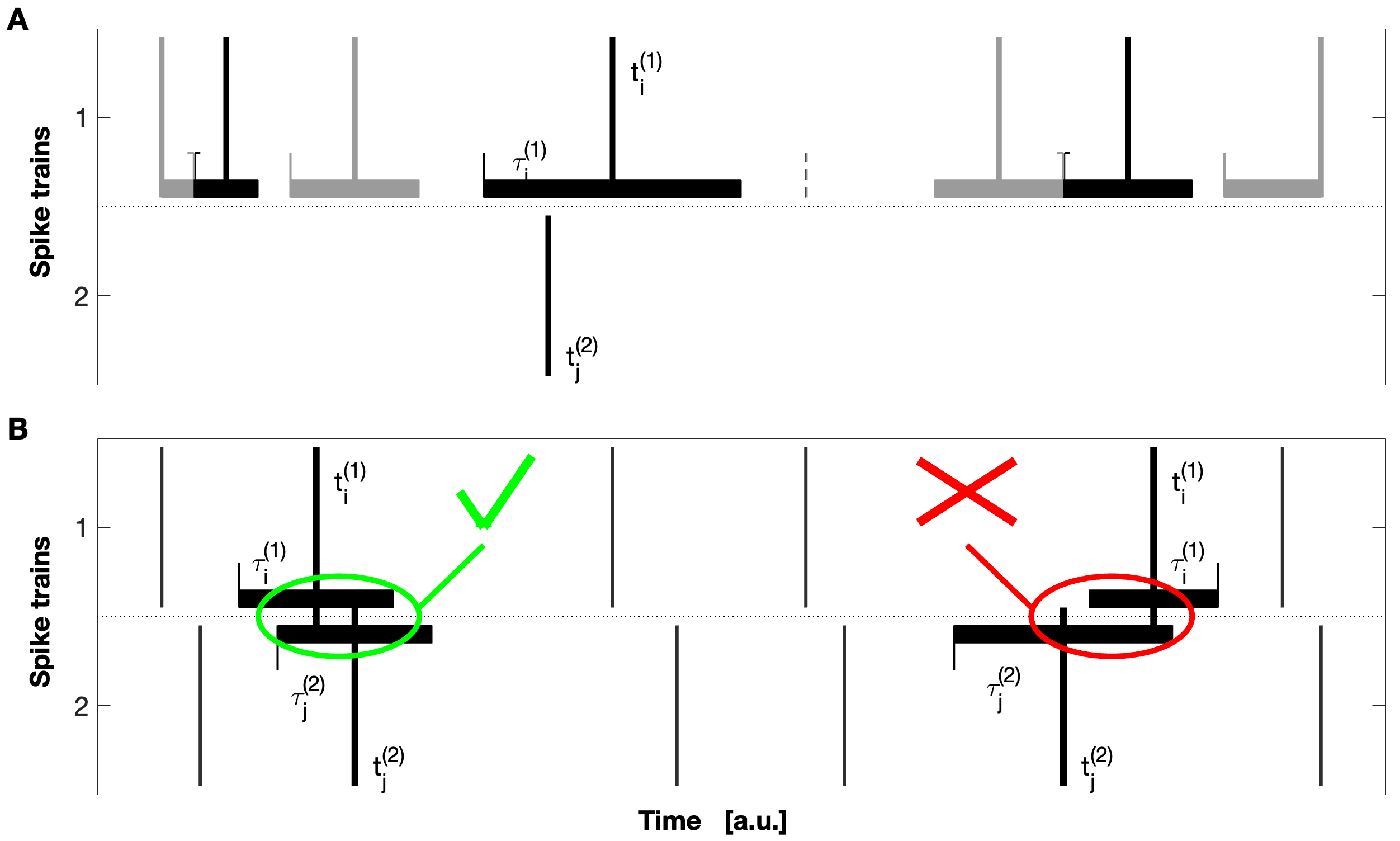}
	\caption{A. Illustration of adaptive coincidence detection.
For clarity spikes and their coincidence windows are shown alternatingly in bright and dark color.
The first step assigns to each spike $t_i^{(1)}$ of the first spike train a potential coincidence window which does not overlap with any other coincidence window:
$\tau_i^{(1)} = \min \{t_{i+1}^{(1)} - t_i^{(1)}, t_i^{(1)} - t_{i-1}^{(1)}\}/2$.
Thus any spike from the second spike train can at most be coincident with one spike from the first spike train.
Small vertical lines mark the times right in the middle between two spikes, and such a line is dashed when it does not mark the edge of a coincidence window.
B. In the same way a coincidence window
$\tau_j^{(2)} = \min \{t_{j+1}^{(2)} - t_j^{(2)}, t_j^{(2)} - t_{j-1}^{(2)}\}/2$
is defined for spike $t_j^{(2)}$ from the second spike train.
For two spikes to be coincident they both have to lie in each other's
coincidence window which means that their absolute time difference has to be smaller than $\tau_{ij}=\min \{\tau_i^{(1)}, \tau_j^{(2)}\}$ (which is equivalent to the definition found in Eq. \ref{Eq:Coincidence-MaxDist}).
For the two spikes on the left side this is the case, whereas the spikes on the right side are not coincident. Thus on the left both coincidence indicators $C_i^{(2,1)}$ and $C_j^{(1,2)}$ equal $1$, while on the right both are $0$. [Modified from \cite{Kreuz17}.]}
	\label{Fig2:Coincidence-Detection}
	\end{center}
\end{figure}

There exists also a variant with a window of fixed size \citep{QuianQuiroga02b}, but the coincidence detection that we use here is scale- and parameter-free since the maximum time lag $\tau^{(m,n)}_{ij}$ up to which two spikes $t_i^{(m)}$ and $t_j^{(n)}$ of spike trains $m, n=1,...,N$ (with $N$ denoting the number of spike trains) are considered to be synchronous is adapted to the local firing rates according to 
\begin{equation} \label{Eq:Coincidence-MaxDist}
 \begin{aligned}
    \tau^{(m,n)}_{ij} = \min \{&t_{i+1}^{(m)} - t_i^{(m)}, t_i^{(m)} - t_{i-1}^{(m)},\\
    &t_{j+1}^{(n)} - t_j^{(n)}, t_j^{(n)} - t_{j-1}^{(n)}\}/2.
 \end{aligned}
\end{equation}

\tcr{We start with a pair of spikes from two different spike trains since this is the minimal case for which there can be a coincidence or not.} Following \cite{Kreuz15}, we \tcr{then} apply the adaptive coincidence criterion in a multivariate context by defining for each spike $i$ of any spike train $n$ and for each other spike train $m$ a coincidence indicator
\begin{equation} \label{Eq:Coincidence-Indicator}
	C_i^{(n,m)}=\begin{cases}
		1 & {\rm if}  \min_j (|t_i^{(n)} - t_j^{(m)}|) < 
		\tau_{ij}^{(n,m)} \cr
		0 & {\rm otherwise.}
	\end{cases}
\end{equation}

Here the minimum function and the ‘$<$’ guarantee an unambiguous spike matching since any spike can at most be coincident with one spike (the nearest one) in the other spike train (Fig. \ref{Fig2:Coincidence-Detection}A). The coincidence indicator $C_i^{(n,m)}$ is either $1$ or $0$ depending on whether the spike $i$ of spike train $n$ is part of a coincidence with a spike of spike train $m$ or not (Fig. \ref{Fig2:Coincidence-Detection}B).

% Mention somewhere:
% For some applications it might be appropriate here to also introduce a maximum coincidence window $\tau_{max}$ \cite{Kreuz17} as a parameter thereby combining the time-scale independent coincident detection with a time-scale dependent upper limit. This way additional knowledge about the data (such as typical signal propagation speed) can be taken into account in order to guarantee that two coincident spikes are really part of the same meaningful event.

\subsection{SPIKE-synchronization, SPIKE-Order and the Synfire Indicator \label{ss:Methods_1-2_Spike-Sync-Order-Synfire-Indicator}}

Applying this adaptive criterion of ‘‘closeness in time" to all spikes of a given spike train set we can obtain an overall measure of spike matching. The resulting SPIKE-synchronization value $C$ (see \ref{App2-SPIKE-synchronization} for the more detailed definitions and equations) quantifies the overall fraction of coincidences \cite{Kreuz15}. While for irregular spiking rather low values are obtained (the expectation value for Poisson spike trains of equal rates is $C = 0.25$ \cite{Mulansky15}), this value increases the more a given spike train set can be separated into complete global events in which all spikes match each other (compare Figs. \ref{Fig1:Overview-Rasterplots}A and \ref{Fig1:Overview-Rasterplots}B whose SPIKE-synchronization values are $0.262$ and $1$, respectively). \tcr{The adaptive nature of the coincidence detection guarantees that this also works for global events that are both of varying duration and irregularly spaced.}

In the next step we also consider the temporal order of the spikes. This SPIKE-Order approach allows to sort the spike trains from leader to follower and to evaluate the consistency of the preferred order via the Synfire Indicator \cite{Kreuz17}. The absolute value of the Synfire Indicator $F$ can never be larger than SPIKE-synchronization $C$ (this along with a full derivation of SPIKE-Order $D$, Spike Train Order $E$ and Synfire Indicator $F$ is detailed in \ref{App3-SPIKE-Order-Synfire-Indicator}). Therefore, $F$-values for irregular spiking must by necessity be close to $0$, while the value for spike train sets with well-defined global events increases with the consistency of the spike order. The sorting of the spike trains can make a big difference: For unsorted spike trains the Synfire Indicator can cover the whole range from $-1$ to $1$, while the optimized Synfire Indicator for sorted spike trains can never be negative and gets the closer to $1$, the more the spike trains resemble a perfect synfire chain.

This transition from irregular spiking (Fig. \ref{Fig1:Overview-Rasterplots}A) via unsorted global events (Fig. \ref{Fig1:Overview-Rasterplots}B) to an almost perfect synfire chain (Fig. \ref{Fig1:Overview-Rasterplots}C) is reflected in the value of the Synfire Indicator $F$ which rises from $-0.018$ (Fig. \ref{Fig1:Overview-Rasterplots}A  before sorting) to $0.071$ (after sorting these spike trains, not shown) via $-0.155$ (Fig. \ref{Fig1:Overview-Rasterplots}B before sorting) and $0.417$ (Fig. \ref{Fig1:Overview-Rasterplots}B after sorting, not shown) to $0.952$ (Fig. \ref{Fig1:Overview-Rasterplots}C, before and after sorting, the spike train order is already optimal). In this article the initial spike trains will always already be sorted from leader to follower.

% XXXXX Big question: After adding noise, should we always first sort the spike trains? Could it be possible that the results would change quite a lot (at least for some of the direct shift methods, e.g., diagonals, also the labelling of the rows could depend on that)? XXXXX YYYYY ideally yes we should sort them, but at the moment we don't have an algorithm for that YYYYY
% XXXXX Expectation value of F for Poisson spike trains ??? Numerically I get a value around 0.0013. XXXXX
% YYYYY should be 0 right? YYYYY XXXXX No, you'll always get statistical deviations from $0$, similar to the random walk where after N steps you end up at a distance ~ sqrt(N) and not at $0$. XXXXX

\subsection{Latency correction} \label{ss:Methods_1-3_Latency-Correction}
% 
% #####################################################################
% ################## Figure 3: Synfire Chain ##########################
% #####################################################################
%
\begin{figure}[!ht]
	\includegraphics[width=\linewidth]{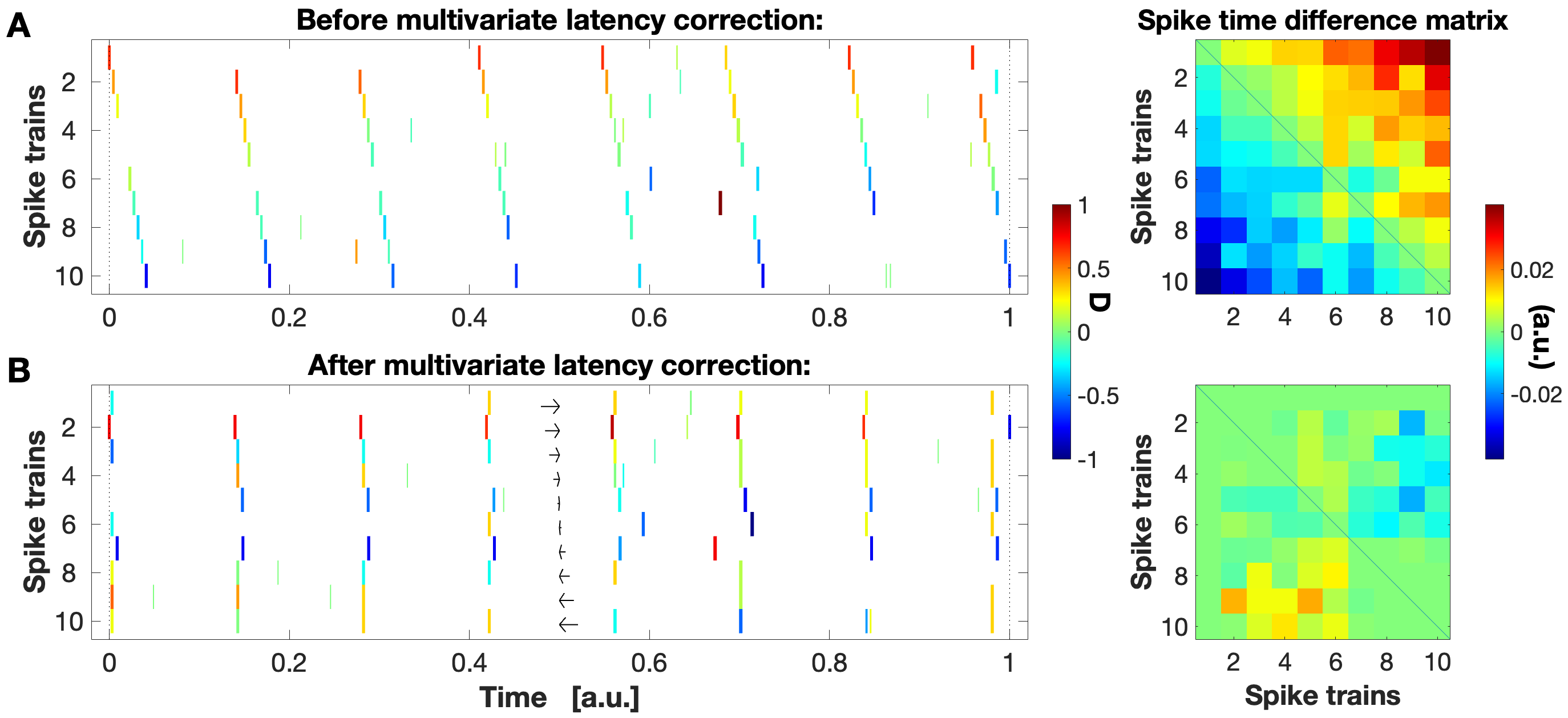}
	\caption{Illustration of a set of artificially created spike trains exhibiting a consistent propagation activity, in this case a regularly spaced synfire chain with some missing spikes, some noisy background spikes and a small amount of jitter. Subplot A is before and subplot B is after the latency correction. The arrows in B indicate for each spike train the shift performed during the latency correction. Spike colors code the anti-symmetric SPIKE-Order $D$ on a scale from $1$ (red, first leader) to $-1$ (blue, last follower). On the right we show the spike time difference matrices. For this simple example the corrected spike trains are almost identical and accordingly the spike time difference matrix (STDM) turns from its rather ordered increase away from the diagonal (since spikes from more separated spike trains exhibit greater time separation) to very low values everywhere.} \label{Fig3:Spike-Time-Difference-Matrix}
\end{figure}

In case one wants to estimate the synchrony within a spike train set, systemic delays such as the ones in the synfire chain of Fig. \ref{Fig1:Overview-Rasterplots}C constitute a hindrance since the time lags in the data lead to a spurious underestimation of synchrony. To account for this, first a multivariate latency correction has to be performed in which the spike trains are realigned (as in Fig. \ref{Fig1:Overview-Rasterplots}D) before the ‘‘true" synchrony can be calculated. An algorithm for such a multivariate latency correction has recently been proposed in \citep{Kreuz22} and since this algorithm plays a crucial part in our approach to deal with overlapping events we here describe it in some detail.

% As has been shown in \citep{Kreuz22}, the effect of the latency correction is more pronounced for spike train sets with high SPIKE-synchronization $C$ and high Synfire Indicator $F$. As we have just seen, these are sparse spike trains with well-defined global events which moreover display a rather high uniformity among events regarding their temporal order, in short, spike trains with a lot of systematic delays to be corrected. 

The first step is to go beyond SPIKE-synchronization and SPIKE-Order and not only look at coincidence and order but also at the actual temporal intervals between matching spikes. Therefore, after having matched the spikes using adaptive coincidence detection (via Eqs. \ref{Eq:Coincidence-MaxDist} and \ref{Eq:Coincidence-Indicator}) we calculate the time difference of the matched pair as:
\begin{equation}
	\delta^{(n,m)}_i = t^{(n)}_i - t^{(m)}_{j'}.
\end{equation}
Here $j'$ identifies the spike in spike-train $m$ that matches spike $i$ in spike train $n$ as
\begin{equation} \label{Eq:Matching-Spike}
	j' = \arg \min_j ( |t_i^{(n)} - t_j^{(m)}| ).
\end{equation}

Finally, averaging over all matches of any pair of spike trains
\begin{equation} \label{Eq:Average-Spike-Time-Difference}
	\delta^{(n,m)} =  \frac{1}{\sum_i C_i^{(n,m)}} \sum_i C_i^{(n,m)} \delta_i^{(n,m)}
\end{equation}
\noindent results in an antisymmetric $N \times N$ matrix called the \emph{spike time difference matrix} (or short STDM) which provides the best estimates of all the pairwise latencies between the spike trains. Similarly, from the same quantities we can define a symmetric \emph{cost matrix} as
\begin{equation} \label{Eq:Cost}
	c^{(n,m)} =  \sqrt{\frac{1}{\sum_i C_i^{(n,m)}} \sum_i C_i^{(n,m)} [\delta_i^{(n,m)}]^2},
\end{equation}
\noindent which, in contrast to Eq. \ref{Eq:Average-Spike-Time-Difference}, guarantees that the value $0$ is obtained if and only if all matched spike pairs are exactly coincident. Therefore, using  the cost matrix $c^{(n,m)}$ instead of the simpler STDM matrix $\delta^{(n,m)}$ (as was done in \cite{Kreuz22}) is an important improvement, at least in the case of more noisy data. The aim of latency correction then becomes to align the spike trains by minimizing the \emph{cost function} $c$, the mean value of the upper right triangular part of the matrix, i.e., all values for which $n < m$:
\begin{equation} \label{Eq:Cost-function}
	c = \frac{2}{N(N-1)} \sum^{N}_{n<m} c^{(n,m)}.
\end{equation}
% = \langle \Delta_< \rangle

On the right hand side of Fig. \ref{Fig3:Spike-Time-Difference-Matrix} we show the STDM for a regularly spaced synfire chain with both missing and extra spikes as well as some jitter, before and after the latency correction. Since for the initial synfire chain the matched spikes of more distant spike trains are further apart, the values in the STDM tend to increase with the distance from the diagonal (which by definition is always $0$). After the latency correction the STDM approaches $0$ everywhere reflecting a much better alignment of the spike trains, but since some jitter remains it does not go all the way to $0$.

In \citep{Kreuz22} two major latency correction algorithms were proposed. In the first algorithm, \textit{direct shift}, only a minimal part of the matrix is used ($N - 1$ values), for example in the \textit{row direct shift} variant the values from one row (typically the first, such that the first spike train is used as reference) or in the \textit{first diagonal direct shift} variant the values from the first upper diagonal (the cumulative difference between neighboring spike trains). The correction is performed and the spike trains are shifted such that the corresponding matrix elements are set to $0$ and the hope is that this way also the other matrix elements are automatically taken care of.

The search space of all possible shifts is infinitely large, therefore the second latency correction algorithm proposed in \citep{Kreuz22} uses the probabilistic heuristic \textit{simulated annealing} which employs an iterative Monte Carlo algorithm to minimize the cost function and find the best alignment of the spike trains. Starting from the cost value of the original spike train difference matrix $c_{start}$, in each iteration a randomly selected spike train is shifted by a random time interval, the spike train difference matrix and the cost function are updated and the shift is accepted whenever the cost function decreases. Unlike greedy algorithms, simulated annealing is able to escape from local minima, since increases of the cost function are allowed with a certain likelihood. The likelihood gets lower and lower as the temperature of the cooling scheme decreases and this iterative random walk continues until the cost function converges towards its minimum cost $c_{end}$.

In both cases the input is a set of $N$ spike trains and in return there are two major outputs: the end cost $c_{end}$ and the shifts $\vec{s} = [s_1;...;s_N]$ performed in order to get there. In \citep{Kreuz22} it was shown that simulated annealing in general achieves better results than direct shifts (e.g., lower end costs). However, the drawback of simulated annealing is that it usually takes several magnitude longer to calculate as the complexity of computation is of the order $N^2$ (whereas both direct shifts consist of just $N - 1$ additions or subtractions).

% #############################################################################
% #################### Section: Methods II ####################################
% #############################################################################
%
\section{Methods II - New approaches to deal with event overlap} \label{s:Methods2}

\begin{figure}[!ht]
	\begin{center}
	\includegraphics[width=\linewidth]{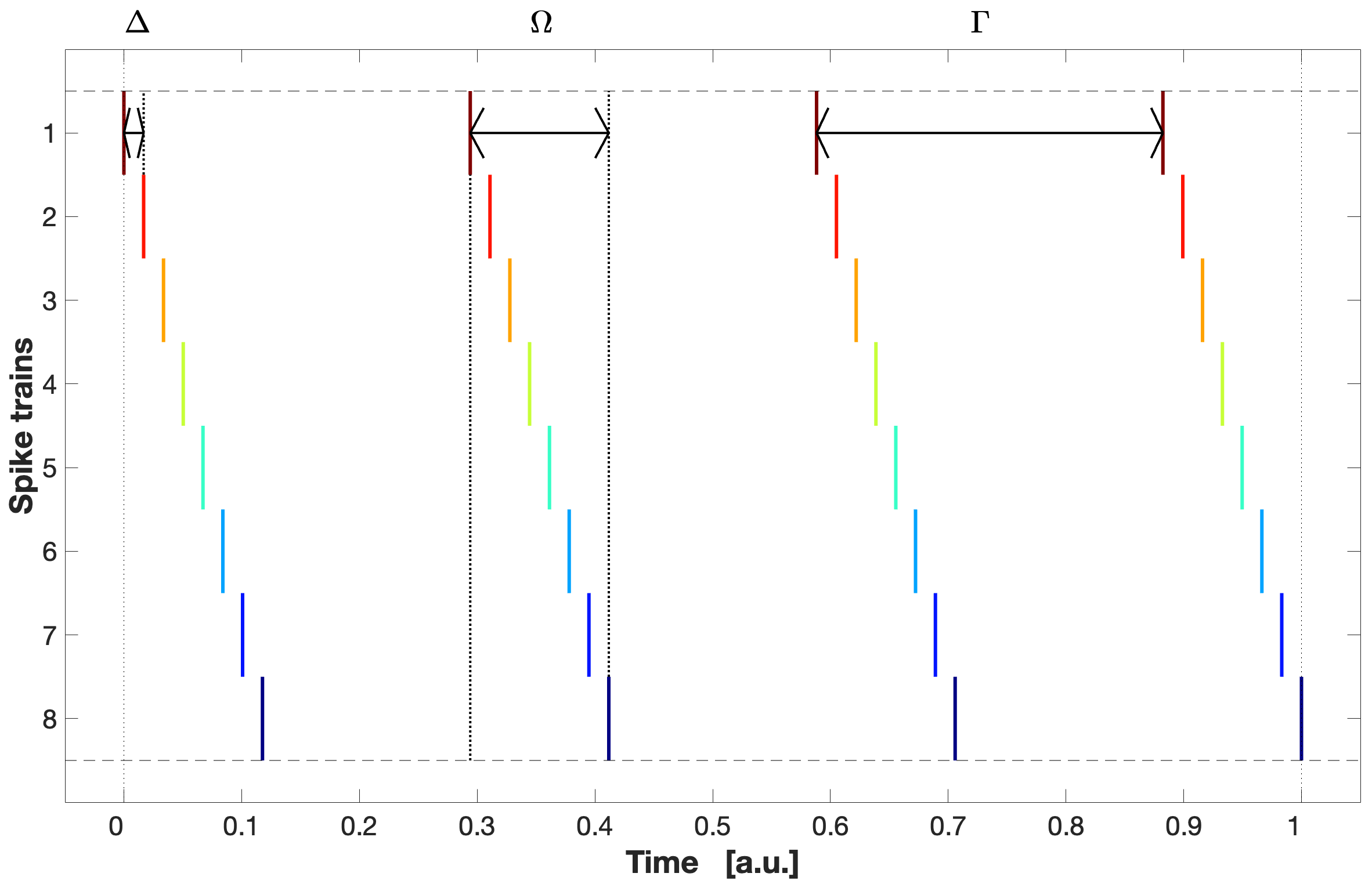}
	\caption{Perfectly clean synfire chain with which we introduce the three quantities that are used in Section \ref{ss:Methods_2-1_Overlap-Theory} to derive the theory of event overlap: the latency $\Delta$ between successive spikes of the same event, the event duration $\Omega$, and the interval $\Gamma$ between successive events.}
	\label{Fig4:Labels}
	\end{center}
\end{figure}

For the sake of simplicity, in Section \ref{ss:Methods_2-1_Overlap-Theory} we illustrate the problem of event overlap on a perfect synfire chain, a succession of complete global events with constant latencies $\Delta$ between successive spikes of the same event (and thus constant event durations $\Omega$) and constant intervals $\Gamma$ between successive events. These three variables are illustrated in Fig. \ref{Fig4:Labels}. However, the new algorithms for latency correction we then propose in Section \ref{ss:Methods_2-2_New-Algorithms} to tackle event overlap are designed to function also on datasets exhibiting higher levels of noise (incomplete events, background noise, jitter etc.). Finally, in Section \ref{ss:Methods_2-3_Relative-Shift-Error} we introduce a new way to quantify the performance of these algorithms when the ground truth is known.
% 
% #########################################################################
% ################## Figure 4: Definition Labels ##########################
% #########################################################################
%

% #############################################################################
% #################### Section: Methods II-1 ##################################
% #############################################################################
%
\subsection{Theory of event overlap} \label{ss:Methods_2-1_Overlap-Theory}

% 
% ############################################################################
% ################## Figure 5: Illustration Overlap ##########################
% ############################################################################
%
\begin{figure*}[th!]
	\centering
	\subfloat[]{\includegraphics[width=\textwidth]{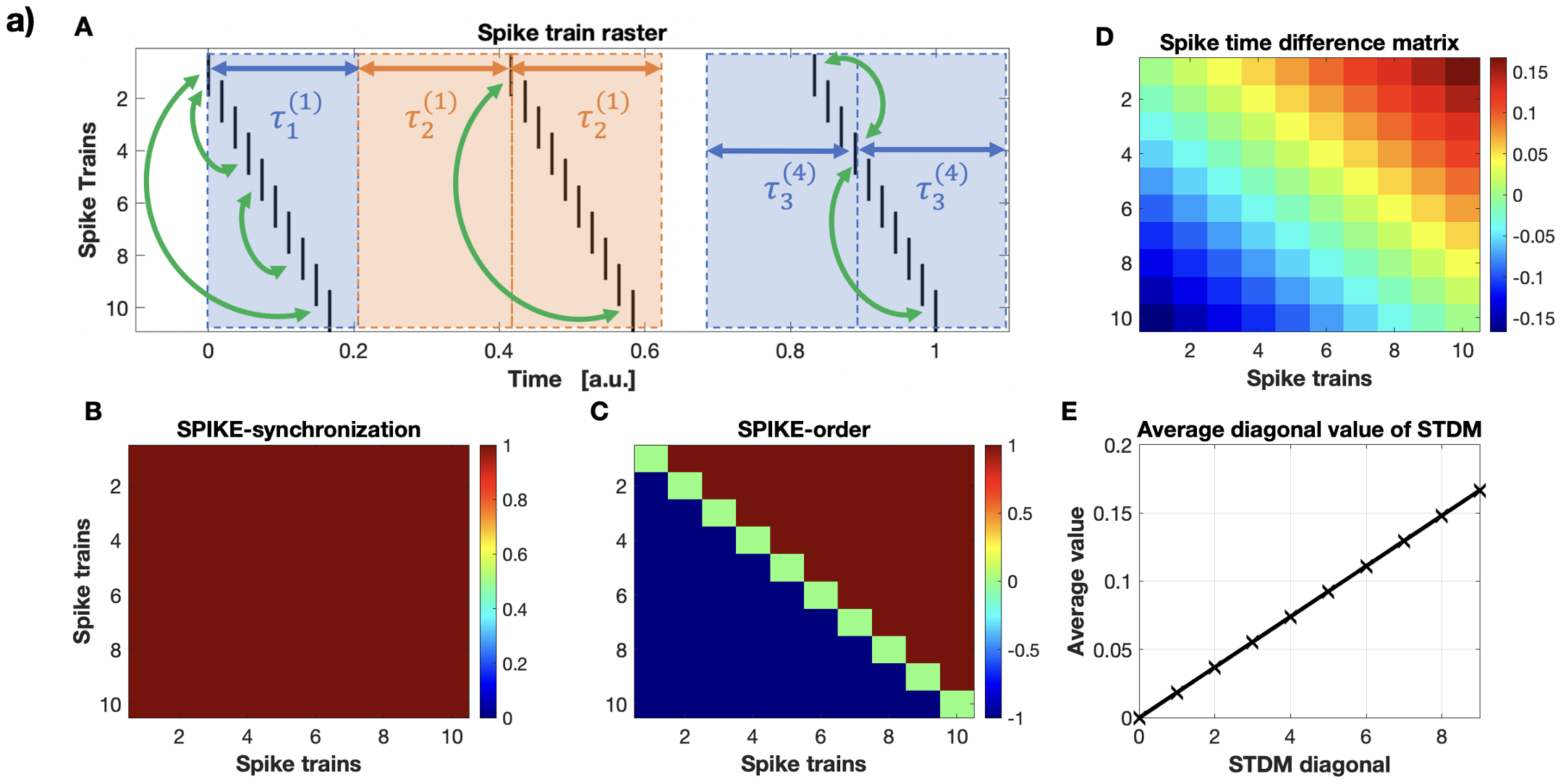}}
	\vspace{-5mm}
	    %\vspace{-17mm}
	{\includegraphics[width=\textwidth]{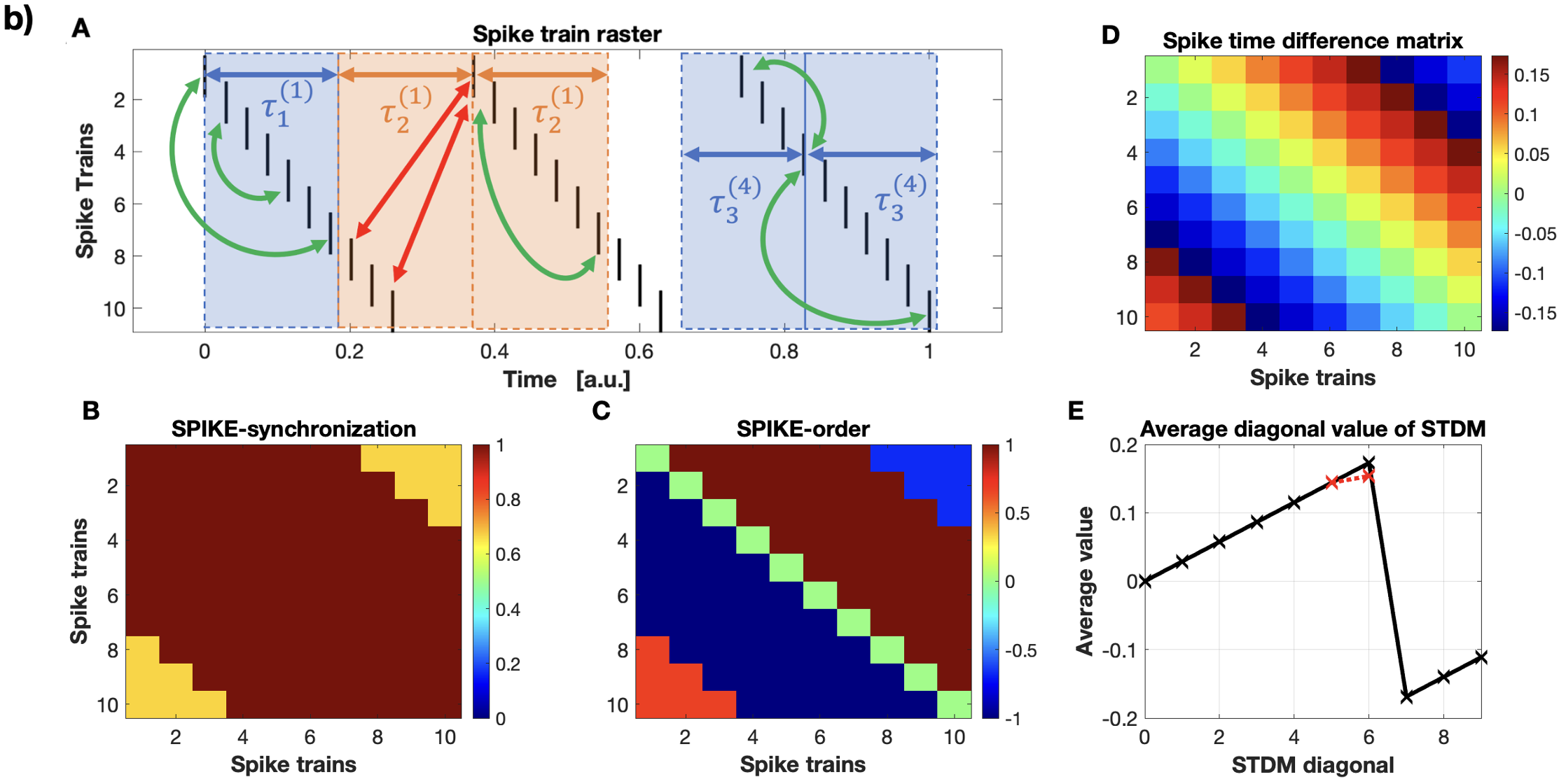}}
	\vspace{-5mm}
	    %\vspace{-17mm}
	\caption{Perfect synfire chain consisting of three global events without (subpanel a) and with (subpanel b) event overlap. Both subpanels follow the same structure:
A. Spike train sets. Alternate coincidence windows of the \emph{i-th} spike in the \emph{(n)-th} spike train $\tau^{(n)}_i$ are indicated in blue and orange. Green (both subplots) and red (subplot b only) arrows mark some exemplary correct and spurious matches, respectively.
B. SPIKE-synchronization and C. SPIKE-Order matrix. At $R = 0.4$, without overlap (a) both matrices exhibit maximal values, whereas in (b) an overlap of $R = 0.7$ leads to spurious matches indicated by deviations from $1$ in the off diagonal corner.
D. Spike time difference matrix (STDM) and E. Average diagonal value of the STDM.
Without overlap (a) values of the STDM increase \tcr{monotonically} with distance from the diagonal, whereas with overlap (b) the maximum is reached at an intermediate diagonal (here the last unaffected diagonal is $\nu = 6$). The short dotted red line in (b) shows an example of how the sixth diagonal could already be affected by the overlap but would still exhibit the maximum average value.}
	\label{Fig5:Overlap-Problem}
\end{figure*}
All the methods described in Section \ref{s:Methods1} work very well as long as the global events are sufficiently separated. Fig. \ref{Fig5:Overlap-Problem}a, subplot A, shows such a perfect synfire chain. In this case each spike of every global event is coincident with all other spikes of the same event and all spikes are in perfectly consistent order. Accordingly, all pairwise values in both the SPIKE-synchronization (subplot B) and the SPIKE-Order matrix (subplot C) yield the maximum value of $1$ and, in consequence, the same holds true for the overall SPIKE-synchronization $C$ (Eq. \ref{Eq:SPIKE-synchronization-C}) and the Synfire Indicator $F$ (Eq. \ref{Eq:Synfire-Indicator-F}). The spike time difference matrix (subplot D, Eq. \ref{Eq:Average-Spike-Time-Difference}) exhibits a monotonous increase with distance from the main diagonal, as quantified by averaging the values of different diagonals of the STDM (subplot E).

However, difficulties arise as soon as coincidence windows (Eq. \ref{Eq:Coincidence-MaxDist}) of neighboring events start to overlap, and these difficulties concern all of SPIKE-synchronization, Synfire Indicator and latency correction. In Fig. \ref{Fig5:Overlap-Problem}b, subplot A, we present an example where the propagation within each event has slowed down so much that the last three spikes from each global event are no longer coincident with the earliest spikes from the same event but rather with the earliest spikes from the next event. These spurious mismatches  lead to imperfect SPIKE-synchronization (subplot B) and inconsistencies in the SPIKE-Order (subplot C), and consequently both overall SPIKE-synchronization $C = 0.956$ and Synfire Indicator $F = 0.778$ are lower than $1$. Finally, the STDM (subplot D) reaches its maximum not at the corner (as in Fig. \ref{Fig5:Overlap-Problem}a) but at an intermediate diagonal (subplot E).

Under normal circumstances the matching derived via the adaptive coincidence detection of Eqs. \ref{Eq:Coincidence-MaxDist} and \ref{Eq:Coincidence-Indicator} would be perfectly reasonable: the spikes are being matched to the ones closest in time, but in this case due to the overlap some of the spikes matched together belong to different events. However, we can often clearly recognise from the rasterplot that the trailing spikes should belong to the earlier event and not to the later one with which they are matched (or we might have external knowledge not available to the impartial algorithm). Therefore, we can consider the mismatches and the resulting reduction in SPIKE-synchronization, Synfire Indicator and spike time difference as spurious. \tcr{Our aim here} is to find ways to ‘‘overwrite" the naive event matching algorithm and correct this.

Comparing Fig. \ref{Fig5:Overlap-Problem}a and Fig. \ref{Fig5:Overlap-Problem}b it becomes immediately clear that the amount of overlap present within a synfire chain depends on the ratio between two basic quantities (see Fig. \ref{Fig4:Labels}): the \emph{event duration} $\Omega$ (defined as the interval between the first and the last spike of the same event) and the \emph{interval between two successive events} $\Gamma$ (time interval between their first spikes):
\begin{equation} \label{Eq:Event_Overlap}
	R = \frac{\Omega}{\Gamma}.
\end{equation}

If for this \emph{event overlap} $R \in \left[ 0,  \infty \right)$ we obtain values $R \geq 0.5$ (i.e., if a global event lasts as least as long as half the distance to the successive event, $\Omega \geq \frac{\Gamma}{2}$), the two events will overlap, as some of the spikes will enter the successive coincidence window of size $\tau = \frac{\Gamma}{2}$ (Eq. \ref{Eq:Coincidence-MaxDist}). Note that the factor $1/2$ comes from Eq. \ref{Eq:Coincidence-MaxDist} and so at $R = 0.5$ it is only the coincidence windows that start to overlap but this already affects event matching. The spikes of the different events themselves start to overlap only at $R = 1$.
% 
% ############################################################################
% ################## Figure 6: Last Unaffected Diagonal ######################
% ############################################################################
%
\begin{figure}[!ht]
	\includegraphics[width=\linewidth]{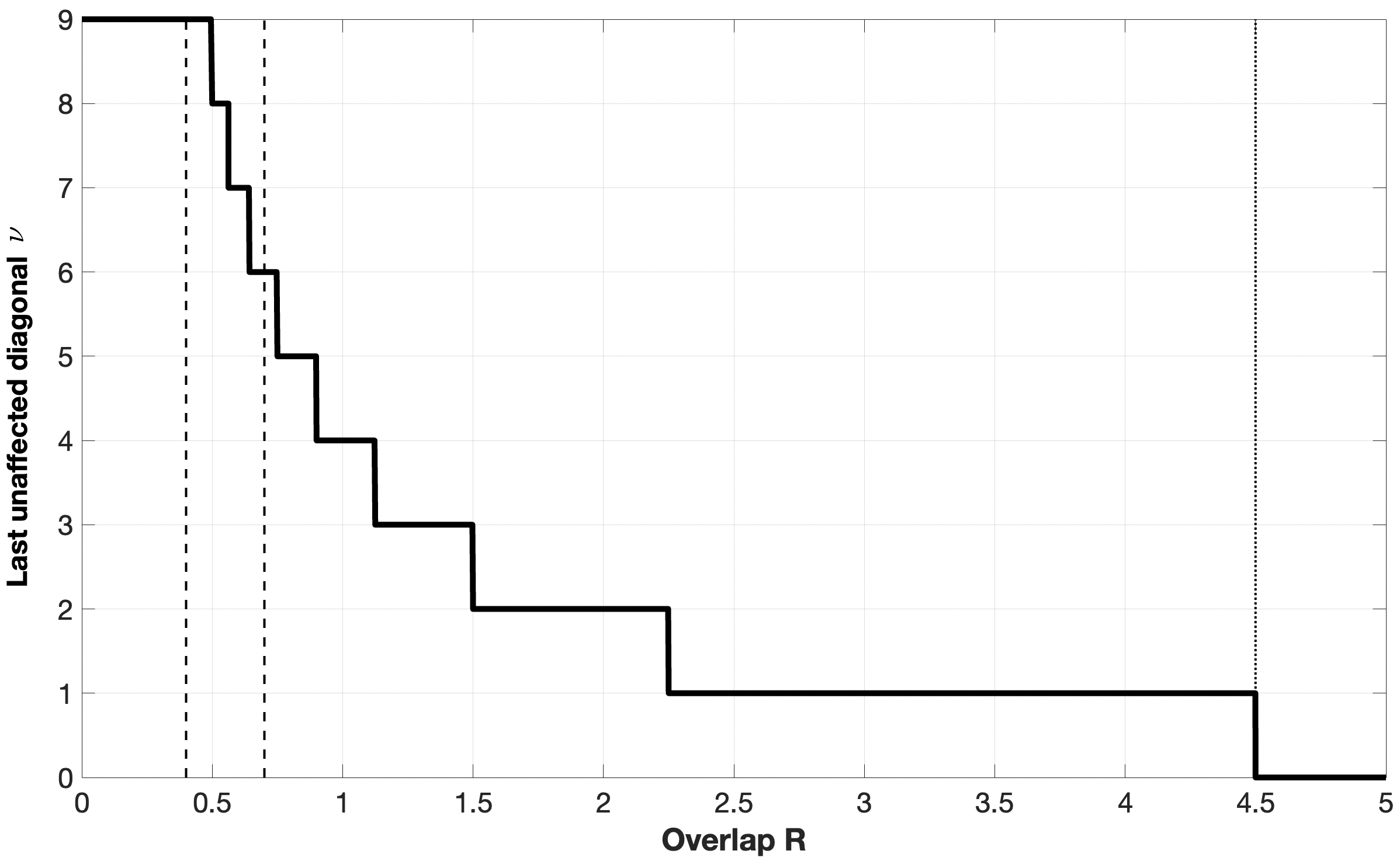}
	\caption{Last unaffected diagonal $\nu$ versus event overlap $R$ for a perfect synfire chain of $N = 10$ spike trains. The two dashed vertical lines correspond to the two examples shown in Fig. \ref{Fig5:Overlap-Problem} ($R = 0.4$ and $R = 0.7$) and the dotted vertical line to the point from which on there are only spurious matches.} \label{Fig6:Last-Unaffected-Diagonal-vs-Overlap}
\end{figure}

Besides $R$, a second useful quantity to describe the amount of overlap is the \emph{last unaffected diagonal $\nu$} in the spike time difference matrix, which can be related to the number of unaffected spikes in a global event, i.e., those that are not (spuriously) matched to the first spike of the following event. The first spike of an event, corresponding to the main (or zeroth) diagonal, can by definition never be affected, while the last unaffected diagonal corresponds to the last spike for which $\nu \cdot \Delta < \frac{\Gamma}{2} = \tau$ still holds. For any synfire chain with overlap (e.g., for $R \geq 0.5$) we immediately obtain
\begin{equation} \label{Eq:last_unaffected_diagonal}
	\nu = \tcr{\textnormal{floor}} \left( \frac{\tau}{\Delta} \right).
\end{equation}
Without overlap, when even the last diagonal is unaffected, we get $\nu = N - 1$ by default.

Eq. \ref{Eq:last_unaffected_diagonal} can be expressed in dependence of the event overlap $R$ by inserting $\Omega = (N-1) \cdot \Delta$ (constant latencies between spikes of the same event), $\Gamma = 2 \tau$ and Eq. \ref{Eq:Event_Overlap} which leads to 
\begin{equation} \label{Eq:last_unaffected_diagonal_vs_event_overlap}
	\nu(R) = \tcr{\textnormal{floor}} \left( \frac{N-1}{2R} \right).
\end{equation}

This relationship between $\nu$ and $R$ is shown in Fig. \ref{Fig6:Last-Unaffected-Diagonal-vs-Overlap} and we can now illustrate it using the two examples of Fig. \ref{Fig5:Overlap-Problem}. In Fig. \ref{Fig5:Overlap-Problem}a the event overlap was set to $R = 0.4$, so there is no overlap and therefore no spurious matches, and thus the diagonal values of the STDM reach their maximum only for $\nu = N-1 = 9$. On the other hand, in Fig. \ref{Fig5:Overlap-Problem}b the event overlap was $R = 0.7$ and there is considerable overlap. The last three spike trains are involved in spurious matches and accordingly the diagonal values of the STDM drop after their maximum at $\nu = 6$. If we go on and further increase $R$ we get more and more overlap until eventually even the spikes from the second spike train start to get involved in spurious event matches. This happens at $\Delta = \frac{\Gamma}{2}$ and thus at $R = \frac{N-1}{2}$ (via Eq. \ref{Eq:Event_Overlap}). From this point on there are no non-spurious spike matchings anymore and any subsequent data analysis based on spike matching (e.g., calculation of $C$ and $F$, latency correction) becomes futile.

It is important to note that the last unaffected diagonal $\nu$ does not always align with the position of the first maximum of the average diagonal values of the STDM. An example is illustrated in subplot E of Fig. \ref{Fig5:Overlap-Problem}b. Initially, since in a normal synfire chain intervals between spikes are longer for spike trains that are further away from each other, these average values always increase. But this continues only as long as the STDM diagonals are unaffected from spurious event matchings. In the case of considerable event overlap, the new matching spikes of the following event are ahead in time instead of earlier in time (as the previous spikes from the same event were) and this leads to a reversal of the sign of these average spike time differences and a sudden drop of the average value. This is the case for the black line in subplot E of Fig. \ref{Fig5:Overlap-Problem}b. However, if overlap occurs gradually and for just some of the matches, this might manifest itself only by a slowing down of the increase (as alluded to by the dashed red line) and in this case the last unaffected diagonal corresponds to the value before the position of the maximum.

In the new algorithms of the next Section \ref{ss:Methods_2-2_New-Algorithms} the last unaffected diagonal $\nu$ will help us via the parameter \textit{stop diagonal} to distinguish the reliable part of the STDM from the unreliable part.

% #############################################################################
% #################### Section: Methods II-2 ##################################
% #############################################################################
%
\subsection{Latency correction with overlap} \label{ss:Methods_2-2_New-Algorithms}

In Section \ref{ss:Methods_1-3_Latency-Correction} we have already summarized our two recently proposed methods of latency correction \citep{Kreuz22}: the first one, \textit{direct shift} is very fast since it extracts the shifts directly from the spike time difference matrix. In particular, it uses only $N - 1$ elements, one for each spike train other than the reference spike train.

For the \textit{row} direct shift variant the identity of the row is a parameter between $1$ and $N$ which defines the reference spike train. The $N - 1$ off-diagonal elements of that row determine directly for the other spike trains the respective shift relative to that reference spike train. Middle rows are less overlap-affected than the outer rows, since they cover less of the STDM corners that are first spoiled by spurious spike matches.

Instead, the second direct shift variant employs only the elements of the \textit{first diagonal}, which is the part of the STDM that is as far away as possible from the overlap-affected corners. The cumulative difference between neighboring spike trains is used such that, for example, the shift between the first and the third spike train is the sum of the shift between the first and the second and the shift between the second and the third spike train.

However, it is clear that both of these direct shift algorithms take into account only a small part of the STDM. In fact, the relative information utilized decreases with the size of the matrix, i.e., with the number of spike trains $N$, as $(N - 1) / \frac{N(N-1)}{2} = 2 / N$. This makes both methods very sensitive to noise in the data: as soon as just one of these $N - 1$ elements is unreliable, the overall error can quickly become very large. This holds particularly true for the first diagonal direct shifts and there especially for the early first diagonal elements since these due to the cumulative nature of the method affect more than just one shift. 
%This works in specific examples (such as a perfect synfire chain) to perfection and in many others comes quite close, but can sometimes also be quite far off.

By contrast, in the second method, \textit{simulated annealing}, the cost function is extracted from the whole STDM which might be a problem in the presence of overlap. Furthermore, the entirety of the latency search space (which in theory is infinitely large) is explored. Both of these facts combine such that without overlap simulated annealing typically yields better results than the two direct shift algorithms but the big drawback is that it also makes this method several magnitudes slower than the direct shifts and for large datasets this might actually be an exclusion criterion.

The most suitable algorithm depends on the situation, and the same holds actually true for datasets with overlap. Therefore, in this Section we will move along two fronts as well. We start by proposing a fast direct shift algorithm termed \textit{Extrapolation} that takes the best of both of the two existing approaches while avoiding the respective drawbacks. In particular, it makes use of a larger part of the spike time difference matrix than the row or first diagonal direct shifts (and is thereby more robust with respect to noise), but at the same time avoids the outer parts of the STDM that are affected by the spurious matches caused by overlap. Subsequently, the same \textit{reduced matrix} approach is also used in the modified simulated annealing algorithm. In both algorithms the extent to which the STDM is used is defined by a parameter, the stop diagonal.

\begin{itemize}

\item Reduced matrix direct shift: Extrapolation

The \emph{Extrapolation} direct shift starts from a reduced part of the spike time difference matrix only and uses the transitivity property to extrapolate the overlap-affected outer parts of the STDM from the unaffected inner parts of the matrix:
\begin{equation}
	\delta^{(n, m)} = \sum_{k=m+1}^{n-1} \Bigl[ \delta^{(n, k)} + \delta^{(k, m)} \Bigr]
\end{equation}
for any $n, m$ with $m - n > d$ where $d$ is the desired \emph{stop diagonal}. We keep the STDM antisymmetric by updating the corresponding elements in the opposite corner as well:
\begin{equation}
	\delta^{(m, n)} = - \delta^{(n, m)}.
\end{equation}

The advantage of having a full STDM (consisting of original and extrapolated parts) for every value of the stop diagonal parameter $d$ is that we can now obtain the shift by simply averaging over the STDM:
\begin{equation}
	s_n =  \frac{1}{N} \sum_{m} \delta^{(m, n)}.
\end{equation}

Since each row/column contains the shifts needed to align with the respective spike train (accordingly the diagonal element is always $0$), averaging over all these shifts immediately results in the invariant shift that uses the median spike train as reference. For more details on this please refer to Section \ref{ss:Methods_2-3_Relative-Shift-Error} and \ref{App4-Relative-shift-error}.

Note that for stop diagonal $d = N-1$ no extrapolation is needed and the average can simply be performed directly over the \textit{full matrix}, whereas for stop diagonal $d = 1$ the Extrapolation direct shift is identical with the \textit{first diagonal} direct shift.

\item Reduced Matrix Simulated Annealing

In case of overlap we can modify simulated annealing along the same lines as we have done for the Extrapolation direct shift. In the original approach the cost function is extracted from the whole upper right triangular part of the spike time difference matrix (see Eq. \ref{Eq:Cost-function} in Section \ref{ss:Methods_1-3_Latency-Correction}). The straightforward correction is to modify the cost function such that only the overlap-unaffected part of the spike time difference matrix is utilized. This is realized by letting the sum in Eq. \ref{Eq:Cost-function} skip all parts beyond the stop diagonal parameter $d$:
\begin{equation} \label{Eq:Reduced-Cost-function}
	c = \frac{2}{(N-d)(N-d-1)} \sum^{N}_{m-n>d} c^{(n,m)}.
\end{equation}
% = \langle \Delta_{d<} \rangle 
%
\end{itemize}
 
Reducing the matrix already leads to a considerable improvement in the performance of the latency correction, but here we propose a way to push the envelope even further. This can be achieved by a novel iterative scheme based on the algorithms introduced above. Each iteration consists of two steps: We start with an initial adaptive coincidence detection to perform a spike matching that relies only on spike pairs that are not affected by overlap (using one of the reduced matrix shifts introduced above). These matches then form the basis of the first latency correction which aims to disentangle the different global events and to eliminate as much overlap as possible. This should considerably improve the spike alignment and thereby increase the performance of the second iteration in which we apply again the two steps of adaptive coincidence detection and latency correction to the newly aligned spike pairs. This time the shift method to be used will depend on whether there is still any overlap present. In both iterations we optimize the performance over the different shift methods. For this we need an appropriate measure of equivalence and similarity between different shifts.

% #############################################################################
% #################### Section: Methods II-3 ##################################
% #############################################################################
%
\subsection{Performance evaluation in controlled data: Relative shift error \label{ss:Methods_2-3_Relative-Shift-Error}}

As introduced in Section \ref{ss:Methods_1-3_Latency-Correction} and extended to the case of global events with overlap in Section \ref{ss:Methods_2-2_New-Algorithms}, every latency correction method outputs \emph{shifts}
$\vec{s} = [s_1;...;s_N]$, an array or vector of time translations to apply to the spike trains in order to achieve the latency correction that minimizes the cost function.

For simulated datasets where latency is artificially introduced (for example in a perfect synfire chain) we know the true latencies and thus the expected shifts $\vec{\sigma}$. In such a case it is therefore possible to define a measure of \emph{relative shift error} $\Sigma$ that compares the shifts $\vec{s}$ delivered by the latency correction algorithm with the correct shifts $\vec{\sigma}$:
\begin{equation} \label{Eq:Relative-Shift-Error}
	\Sigma  = \frac{\lVert \vec{\sigma} - \vec{s} \rVert}{\lVert  \vec{\sigma}  \rVert}.
\end{equation}

This measure is useful to compare the performance of the different shift algorithms not only for clean synfire chains but in a statistical sense also for noisy data where we can average the error over many realizations of the noise. The expectation value of the error might be larger than $0$ but we are still able to select the shift algorithm that results in the smallest error. Before the latency correction the initial shift is zero and we get $\Sigma = 1$ by definition.
% 
% ###################################################################
% ################## Figure 7: Definition Norm ######################
% ###################################################################
%
\begin{figure}[!ht]
	\includegraphics[width=\linewidth]{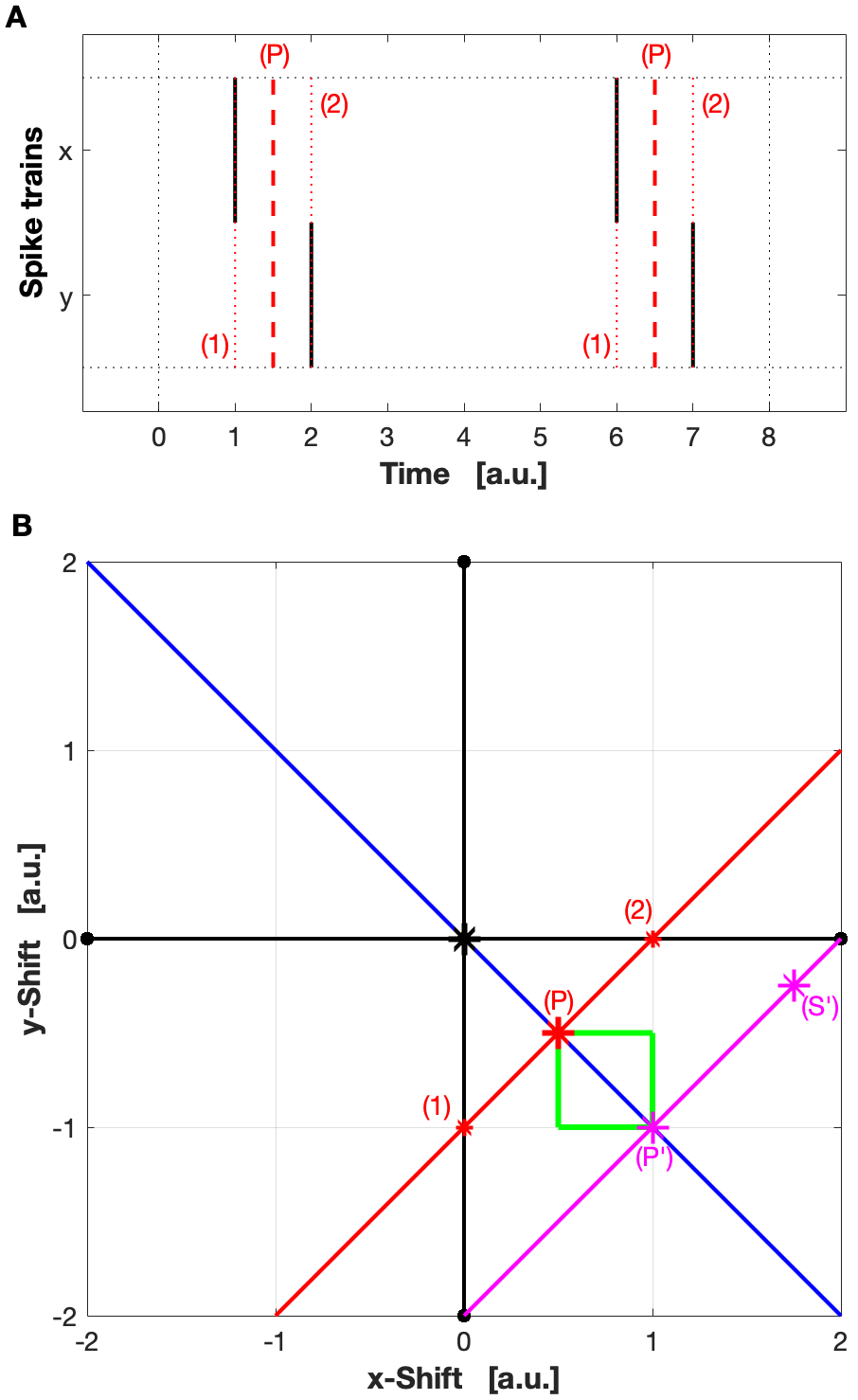}
	\caption{Shift error calculation in simulated data with known ground truth.
	A. Minimal example of a perfect synfire chain with just $N = 2$ spike trains and $2$ spikes each. The second spike train exhibits a latency of $1$ (in arbitrary units) with respect to the first. The three labels (1), (2), and (P) mark the reference times of the three correct (red) shifts indicated in the subplot below.	
	B. A shift of $[0, -1]$ yields a perfect latency correction with the first spike train as reference (1), a shift of $[1, 0]$ with the second (2). However, there are infinitely many shifts that lead to identical spike trains. In the $N$-dimensional shift state space all these shifts align along a line parallel to the identity vector (here the red line). When a latency correction algorithm returns a suggested shift (let us say $[1.75, -0.25]$ (S')) we would like to compare it with the correct shift in a way that is invariant of the reference used. A straightforward way to do so is to project both the correct (red) and the suggested (pink) line onto the $(N - 1)$-dimensional subspace orthogonal to the identity vector (its kernel or null space, here the blue line), thereby minimizing their distance to the origin, and then calculate the distance between them on this subspace. In our example this yields the shifts $[0.5, -0.5]$ (P for projection) and $[1, -1]$ (P'), respectively, and both their absolute (using the taxicab norm, marked in green) and their relative distance is $1$.} \label{Fig7:Relative-Shift-Error}
\end{figure}

The cost function is only dependent on relative time shifts between the spikes and the absolute positions of the spikes do not matter. This means that the cost function and the corresponding shifts to get there are invariant under homogeneous time translation of the whole spike train set. Thus, for each spike train set, there are infinitely many shifts that lead to the same minimal cost function. For this reason in Eq. \ref{Eq:Relative-Shift-Error} we need to use a norm that is invariant under homogeneous time translation (which includes being invariant to which spike train is used as reference, if any).  

As Fig. \ref{Fig7:Relative-Shift-Error} illustrates on the simplest possible synfire chain with just $2$ spike trains and only $2$ spikes each (Fig. \ref{Fig7:Relative-Shift-Error}A), all the shifts resulting in the same relative spike time differences among the spike trains align in the $N$-dimensional shift state space along rays parallel to the identity vector (Fig. \ref{Fig7:Relative-Shift-Error}B). Accordingly, time translation invariance can be accomplished by projecting both the shifts obtained by the latency correction algorithm and the correct shifts onto the $N-1$ dimensional subspace orthogonal to the identity vector. The projection chooses among all equivalent shifts the one that minimizes the distance to the origin, which means that the reference is chosen such that the overall amount of shifts is minimized. 

What remains left to do is to choose a norm. If the $L^2$ (or Euclidean) norm is used the reference is the overall mean of the matched spike pairs, while for the $L^1$ (or Taxicab/Manhattan) norm it is the median (see \ref{App4-Relative-shift-error}) \cite{cormen1990introduction}. Accordingly, to reach the minimum we have to subtract either the mean value (for the $L^2$-norm) or the median value (for the $L^1$-norm) of the shifts. In this article we choose the taxicab metric as it presents a linear relation between the values of latency among spike trains, returns relative shift errors which are less sensitive to outliers in some of the shifts, and can be more intuitively understood.

% #############################################################################
% #################### Section: Results #######################################
% #############################################################################
%
\section{Results} \label{s:Results}

In Section \ref{ss:Results-Sim} we compare the performances of the existing and the newly proposed methods for latency correction in a controlled setting using simulated data that cover the whole range from a perfect synfire chain to pure Poisson spike trains combined with various levels of overlap. Subsequently, in Section \ref{ss:Results-Exp} we apply the algorithm to the neurophysiological datasets described in Section \ref{s:Data}, single-unit recordings from two medial superior olive (MSO) neurons of an anaesthetised gerbil.
% 
% ###################################################################
% ################## Figure 8: Example Simulated Data ###############
% ###################################################################
%
\begin{figure*}[!thb]
    \centering
    \begin{subfigure}[b]{\textwidth}
        \centering
        \includegraphics[width=\textwidth]{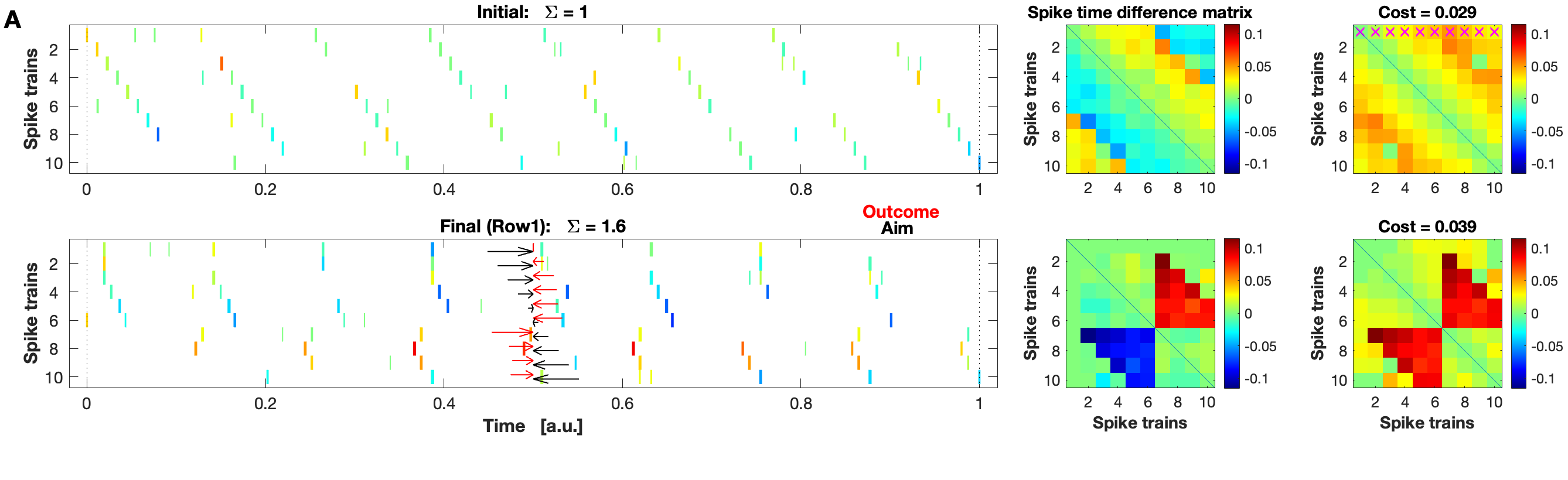}
    \end{subfigure}
    %\vspace{3cm}
    \begin{subfigure}[b]{\textwidth}  
        \centering 
        \includegraphics[width=\textwidth]{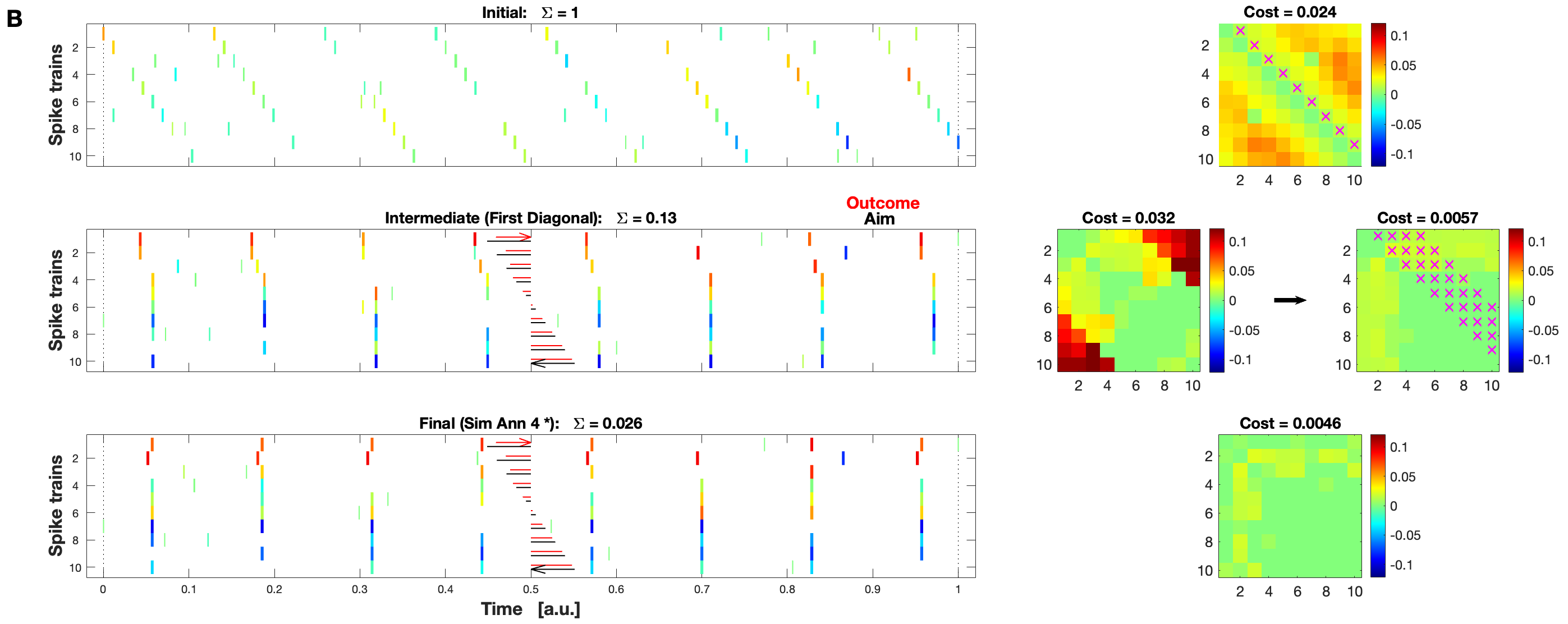}
    \end{subfigure}
    %\vspace{-.5cm}
    \caption{Two different latency corrections applied to a simulated synfire chain with overlapping incomplete events and some background spikes (same overlap $R = 0.8$ and same mixing value $x = 0.2$). In subpanel A we show both the spike time difference and the cost matrices, whereas in subpanel B we restrict ourselves to the cost matrices.
A. Direct shift based on spike time difference values of the \textit{first row} only (marked by red crosses in the cost matrix on the far right). This direct shift includes erroneous information from the outer parts of the spike time difference matrix that are affected by spurious spike matches. It thus falsely tries to align pairs of spikes that are actually mismatched and this results in the breaking up of one global event (as previously seen in Fig. \ref{Fig1:Overview-Rasterplots}F) and accordingly in a very high relative shift error of $\Sigma = 1.6$.
B. Iterative scheme composed of a \textit{first diagonal direct shift} (which thus largely avoids the spurious spike matches in the corner regions of the STDM) and a \textit{reduced matrix simulated annealing} which aims to minimize the cost values up to the fourth diagonal only (again marked by red crosses). The first iteration seems to actually worsen (increase) the cost value (since it is still based on spurious spike matches), but once the rematching is performed (indicated by the black arrow on the right), the improvement is considerable. The simulated annealing of the second iteration still leads to a further improvement which can be seen both in the relative shift errors $\Sigma$ on the left and the cost values on the right.
In both subpanels A and B, for comparison, the shifts that would eliminate the systematic delays of the synfire chain are marked in the later spike train plots by black arrows (Aim), while the shifts obtained by the diverse latency correction algorithms are indicated by red arrows (Outcome).} 
\label{fig:Fig8-Simulated-Examples}
\end{figure*}

% #############################################################################
% #################### Subsection: Results:Sim ################################
% #############################################################################
%
\subsection{Simulated data} \label{ss:Results-Sim}

To test and compare the different shift methods on controlled data with known ground truth, we construct synfire chains for which the overlap $R$ between successive events is varied from $R = 0.4$ (no overlap) to $R = 3$ in steps of $0.2$. As second parameter intended to characterize noise in a rather general way, we also use a mixing parameter $x$, first introduced in \cite{Kreuz22}, to interpolate between the two extremes of perfect synfire chain ($x = 0$) and pure Poisson spike trains ($x = 1$). This mixing parameter is increased from $0$ to $1$ in steps of $0.1$ and for any given $x$ we select a fraction $x$ of spikes from a perfect synfire chain (with $N = 10$ spike trains containing $M_n = 8$ spikes each) and a fraction $1-x$ from pure Poisson spike trains (with an expectation value of $\langle M_n \rangle = 8$ spikes). This way we create a superposition of a synfire chain and a set of Poisson spike trains with the relative contribution determined by the mixing parameter. For every one of these $14 \times 11 = 154$ parameter combinations we average over $N = 100$ spike train sets so that we get a total of $15400$ realizations.

In Fig. \ref{fig:Fig8-Simulated-Examples} we show two exemplary latency corrections to slightly noisy synfire chains (mixing $x = 0.2$) with overlapping events ($R = 0.8$). In the first example (Fig. \ref{fig:Fig8-Simulated-Examples}A) we use the previously proposed row direct shift (\cite{Kreuz22}, see Section \ref{ss:Methods_1-3_Latency-Correction}) where the shift relies on the elements of the first row only and thus also takes into account some of the overlap-affected values in the off-diagonal corner of the STDM which leads to an excessively high relative shift error of $\Sigma = 1.6$ (as defined in Section \ref{ss:Methods_2-3_Relative-Shift-Error}). The second example in Fig. \ref{fig:Fig8-Simulated-Examples}B uses the iterative scheme described at the end of Section \ref{ss:Methods_2-2_New-Algorithms}. In the initial iteration a \textit{first diagonal direct shift} (Section \ref{ss:Methods_1-3_Latency-Correction}) is used to eliminate the influence of the spurious spike matches caused by overlap in the outer STDM elements. While it results in a considerable improvement of the relative shift errors (which decreases from $1$ to $0.13$), this step actually leads to an increase in the cost value (from $0.024$ to $0.032$) but this is only due to the fact that the cost value is still based on spurious spike matches. Once the spikes that are now much better aligned within real global events are rematched, the true cost value turns out to be as low as $0.0057$. In the second iteration we now employ \textit{reduced matrix simulated annealing} with stop diagonal $\nu = 4$ (Section \ref{ss:Methods_2-2_New-Algorithms}) to decrease the cost value further to a final value of $0.0046$ (while at the same time decreasing the relative shift error even more to $0.026$).

% XXXXX Comment somewhere on relative usefulness of the two different indicators (shift error and cost) in case of overlap and dependent on whether a true solution is known or not. Also in the latter case maybe always just use first diagonal in the first iteration, then do rematching, and then start using cost as a meaningful measure (since shift error is not available). Test this on Jason's data (where we actually are able to validate this) XXXXX

% ###################################################################
% ############## Figure 9: Simulated Data - 3D-Results #############
% ###################################################################
%
\begin{figure*}[!thb]
    \includegraphics[width=\textwidth]{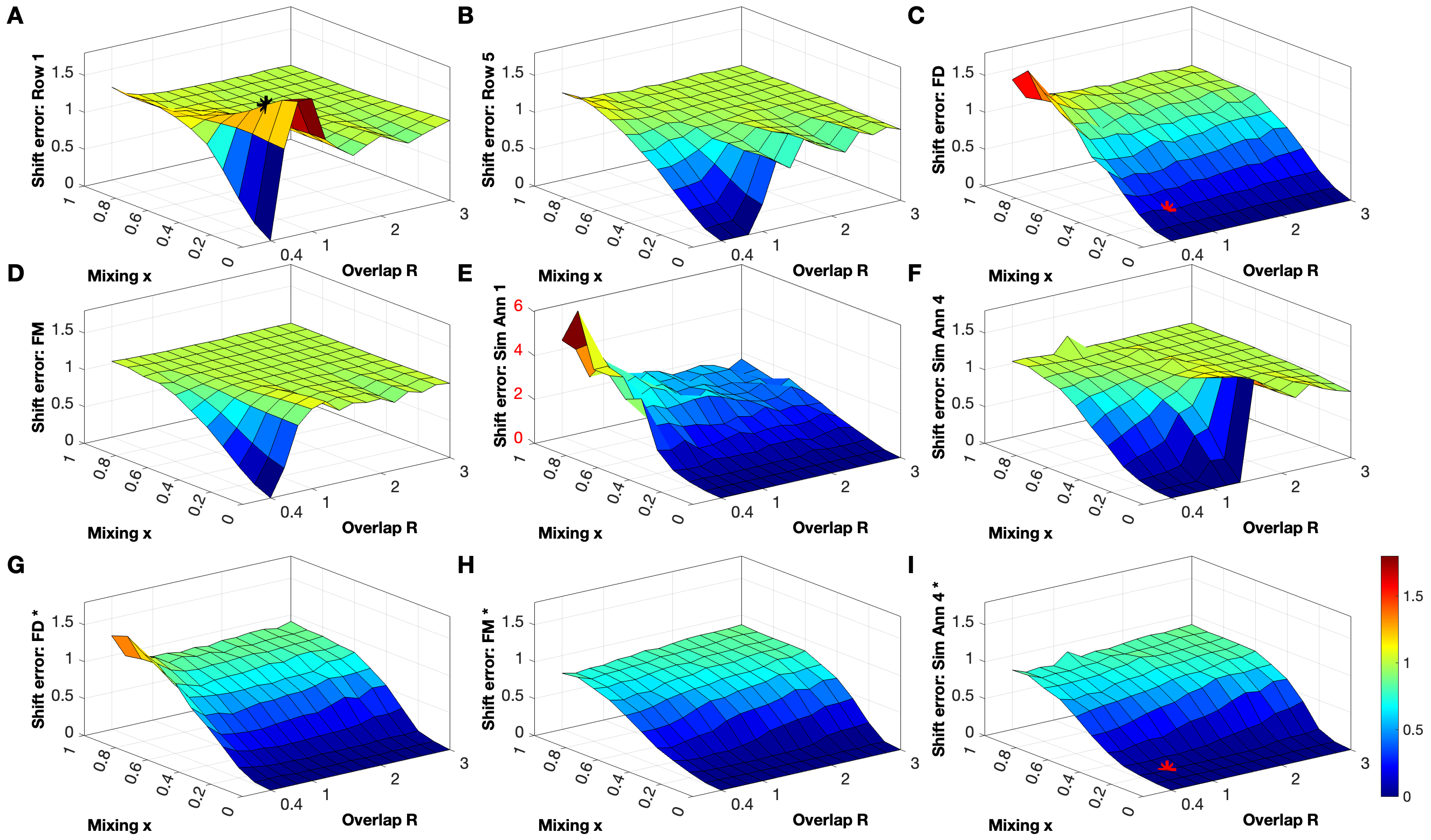}
    \caption{Performance comparison of different latency correction methods on simulated synfire chains with increasing overlap and additional noise (all values are averaged over $100$ noise realizations): 3D-plots of relative shift error $\Sigma$ versus overlap $R$ and mixing parameter $x$. The first two subplots depict the results of two row direct shifts, one outer row (A) whose relative shift error is immediately affected by overlap and one inner row (B) that still exhibits $0$ relative shift errors at least for smaller amounts of overlap. In contrast, the first diagonal (FD) direct shift in subplot C does not show any overlap effects but is instead sensitive to large amounts of noise (in particular, when the overlap is either not present or very small). The opposite (less effect of noise, but high sensitivity to overlap) is true when the direct shift is carried out based on the full matrix (FM) in subplot D and the overall performance is better than for the outer row but worse than the one of the inner row. Regarding reduced matrix simulated annealings, for stop diagonal $\nu = 1$ (E) the noise effect is even worse than for the first diagonal shift (note the different z-scale of this subplot, marked in red). Instead, for stop diagonal $\nu = 4$ (F) the robustness to noise is improved whereas the sensitivity to overlap is worse. Finally, the last row (G-I) shows three exemplary results for the second iteration (again after first diagonal in the first iteration) for which results have improved in two respects: not only have the first latency correction and the subsequent rematching eliminated the overlap effect across the board, but also the absolute shift errors are consistently lower than for the first iteration. The location of the two examples shown in Fig. \ref{fig:Fig8-Simulated-Examples} are marked by a black asterisk in subplot A and a red asterisk in subplots C and I (one for each iteration).}
\label{fig:Fig9-Simulated-3D-Plots}
\end{figure*}

Next, in Fig. \ref{fig:Fig9-Simulated-3D-Plots} we look in a more systematic way at the performance of the different shift methods in dealing with overlapping events and additional mixing noise. We start with two row direct shifts already proposed in \citep{Kreuz22}, one based on an outer and one based on an inner row. Since the first row (Fig. \ref{fig:Fig9-Simulated-3D-Plots}A) covers the off-diagonal corner, it exhibits $0$ error only for the perfectly noise- and overlap-free case. As soon as either overlap or noise is added, this method is immediately affected and its error increases rapidly. In fact, the worst values are obtained for little noise and small overlap when the spurious matches result in one global event breaking up in two (as exemplarily shown in Fig. \ref{fig:Fig8-Simulated-Examples}A). For very high values of the mixing parameter, there is no synfire chain left to correct and accordingly the shift errors plateaus at a rather large value which is independent of the overlap. If instead of the first we use the middle row $5$ (Fig. \ref{fig:Fig9-Simulated-3D-Plots}B), the effect of the overlap is much less pronounced and the error in the noise-free case remains $0$ for very small overlaps and then increases more slowly. On the other hand, the first diagonal direct shift (Fig. \ref{fig:Fig9-Simulated-3D-Plots}C) does not show any overlap effects at all (in fact, for mixing $x = 0$ its shift error stays at $0$ for the whole overlap range), but instead it is very sensitive to large amounts of noise. In particular, for small overlap values a bump is caused by occasional cases where a pair of spike trains does not have any spike pairs which results in a $0$-shift which then in turn leads to a systematic shift error. 

The robustness to noise is much better when all elements of the spike time difference matrix are taken into account (Fig. \ref{fig:Fig9-Simulated-3D-Plots}D), but this case again exhibits a high sensitivity to overlap (somewhat in between the two row direct shifts). Next, we turn our attention to two reduced matrix simulated annealings. For stop diagonal $\nu = 1$ (Fig. \ref{fig:Fig9-Simulated-3D-Plots}E) we observe the same noise effect as for the first diagonal shift (Fig. \ref{fig:Fig9-Simulated-3D-Plots}C), but now it is even worse because in this case shifts between two spike trains without any matches will always be accepted since the cost does not change. Indeed, in order to keep this effect within reasonable bounds we actually added an extra penalty to the cost function that severely punishes (and thus prevents) cases where a spike train would be moved outside of the bounds of the other spike trains. For stop diagonal $\nu = 4$ (Fig. \ref{fig:Fig9-Simulated-3D-Plots}F) the noise effect is much less prominent but on the other hand there is again more sensitivity to overlap.

So far, we have only compared results of the first step of our iterative scheme. When we now look at the relative shift errors obtained after the second iteration (Fig. \ref{fig:Fig9-Simulated-3D-Plots}G-I) we find considerable improvements (apart from the persistent minor bump for the first diagonal direct shift in Fig. \ref{fig:Fig9-Simulated-3D-Plots}G, which is again caused by occasional spike train pairs with no spike matches at all, a case where nothing can be done): The first latency correction using the first diagonal method and the subsequent rematching have in all cases eliminated the overlap effect (without mixing all values remain $0$ for the entire overlap axis). More importantly, the absolute shift errors are consistently lower than for the first iteration. The only effect remaining is an increase with noise but even this increase is now less steep.

% ###################################################################
% ######### Figure 10: Simulated Data - Overall Comparison ###########
% ###################################################################
%
\begin{figure*}[!thb]
    \includegraphics[width=\textwidth]{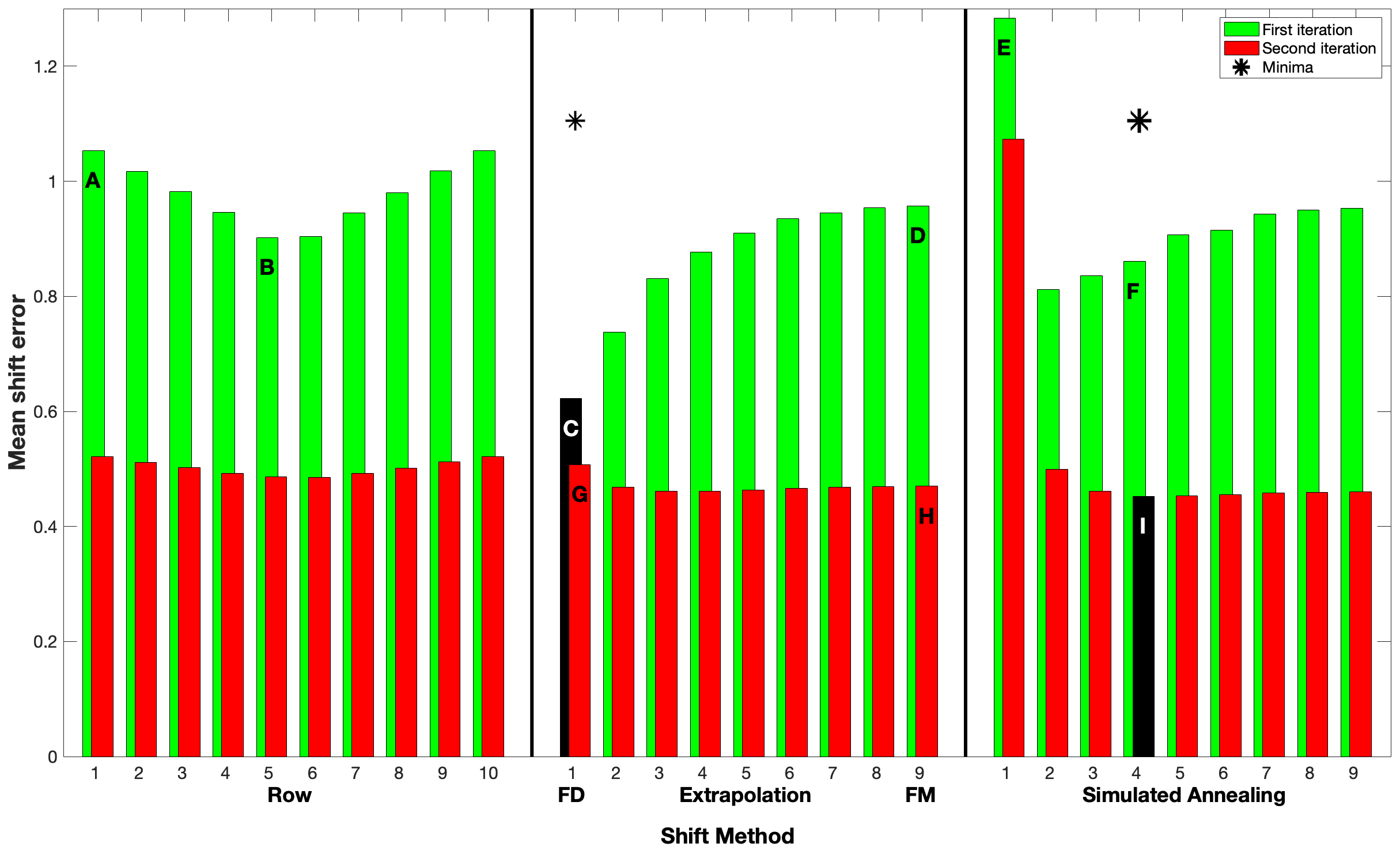}
    \caption{Comparison of different latency correction algorithms both for the first and the second iteration of our iterative scheme: Overall performance averaged over all overlap $R$ and mixing parameter $x$ values (and over $100$ realizations each) for $N = 10$ spike trains. The shift methods depicted in Fig. \ref{fig:Fig9-Simulated-3D-Plots} are indicated by the label of the respective subplot of that Figure. The best performers of the first iteration (first diagonal direct shift) and the second iteration (simulated annealing with stop diagonal $\nu = 4$) are marked in black color as well as with black asterisks. There are two major takeaways: The first iteration shows a high correlation between the use of spurious information from the overlap-affected off-diagonal corners and the shift errors obtained, and the second iteration consistently improves quite a lot on the first iteration. FD = First Diagonal. FM = Full Matrix.}
\label{fig:Fig10-Simulations-Overall-Comparison}
\end{figure*}

The overall comparison of all the different shift methods for both iterations and averaged over all $100$ realizations of all $154$ overlap and mixing values is shown in Fig. \ref{fig:Fig10-Simulations-Overall-Comparison}. The first thing to notice is that the relative shift errors of the first iteration show a clear dependence on how much the elements of the reduced spike time difference matrix (only the ones that are really included by the respective shift method) cover the overlap-affected off-diagonal corners. For example, performance is worse for outer rows than for inner rows, whereas for the reduced matrix methods it increases \tcr{monotonically} from the lowest stop diagonals towards the full matrix. The only exception to this rule is the simulated annealing effect for stop diagonal $\nu = 1$ which is, as already mentioned before, caused by occasional spike train pairs without any spike matches.

The overall best performer of the first iteration is the \textit{first diagonal direct shift} (with an average  minimal shift error of $\Sigma = 0.62$) and this is the method that forms the basis for the rematching carried out ahead of the second iteration. Note that this is a bit of a practical oversimplification. As can be seen in Fig. \ref{fig:Fig9-Simulated-3D-Plots}, different methods have weaknesses at different corners of the noisiness-overlap plane. In general, there is always a tradeoff between noisiness and overlap and the first diagonal direct shift is on one end of that tradeoff. Due to the small number of STDM elements it considers ($N-1$) it is one of the methods that is most sensitive to noise. But since the first diagonal is furthest away from the overlap-affected corners of the STDM, it is the last method that keeps working reasonably well even for quite large amounts of overlap. Starting from that innermost first diagonal, adding more and more diagonals increases robustness to noise but also potential exposure to overlap. It is clear that among the $15400$ datasets analyzed here, there are some (in particular at large amounts of noise and low overlap) for which the minimum error is obtained for a stop diagonal more outwards than the first. However, for the sake of simplicity and convenience, in this large-scale analysis here we refrain from individual optimization in the first iteration and instead choose the overall best performer, the first diagonal, consistently. Within the tradeoff noisiness versus overlap this means erring on the side of least exposure to overlap. \footnote{Later, in the experimental Section \ref{ss:Results-Exp} where we have only $20$ datasets we will do the individual optimization. For example, in the dataset depicted in Fig. \ref{fig:Fig12-Experimental-Data-Latency-Correction-Example}, the optimal stop diagonal after the first iteration happens to be $15$.}

For this second iteration the dependence on overlap with overlap-affected regions of the STDM is largely gone (the only issue remaining is a small dip for the inner row direct shifts). But most importantly, there is a big improvement from the first to the second iteration. Even the highest shift error of the second iteration (again apart from the first diagonal simulated annealing) is considerably smaller than the best shift error of the first iteration. The overall minimal shift error $\Sigma = 0.45$ is obtained for the first diagonal direct shift followed by simulated annealing with stop diagonal $\nu = 4$. However, it should be noted that the Extrapolation direct shift achieves almost the same performance as the much slower simulated annealing (compare second and third block in Fig. \ref{fig:Fig10-Simulations-Overall-Comparison}), a fact that becomes particularly important for the very large datasets that we encounter in the next Section. 

% XXXXX Maybe comment on dependence on mixing and overlap, e.g., first diagonal wins the first iteration somehow by construction (since the average is done for a large range of overlaps including very large ones, for smaller overlap methods with larger stop diagonals would probably outperform the first diagonal). Also these are the average values of 100 realizations. There might be individual realizations where First Diag does not win the first iteration and the second iteration is based on a new spike matching following another shift method. XXXXX

% #############################################################################
% #################### Subsection: Results:Exp ################################
% #############################################################################
%
\subsection{Experimental data} \label{ss:Results-Exp}

% ###################################################################
% ############## Figure 11: Jason - Spike Matches ###################
% ###################################################################
%
\begin{figure}[!thb]
	%\centering % Center the figure on the page
	\begin{subfigure}{0.485\textwidth}
		\includegraphics[width=\linewidth]{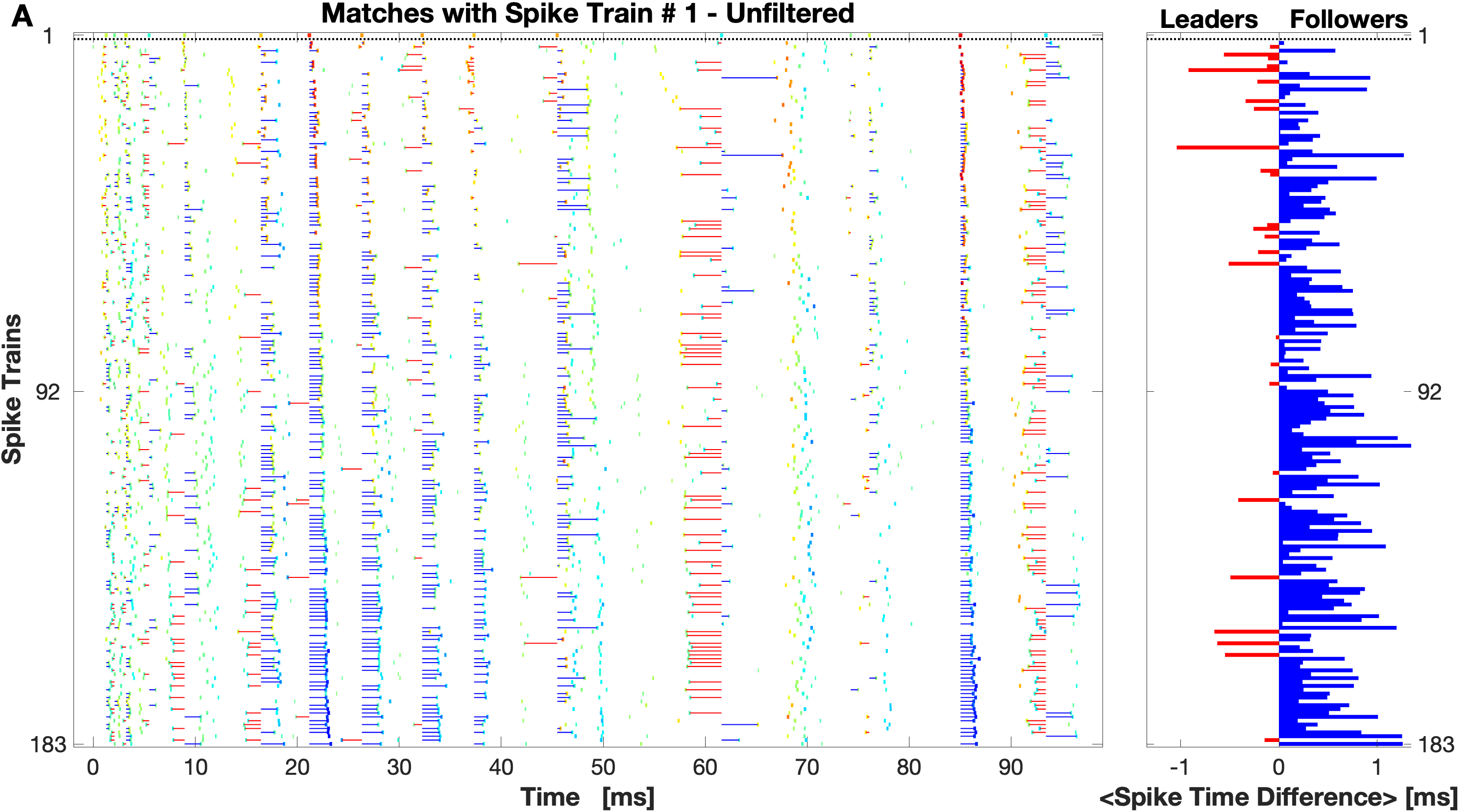}
    \vspace{-1mm}
	\end{subfigure}
	\begin{subfigure}{0.485\textwidth}
		\includegraphics[width=\linewidth]{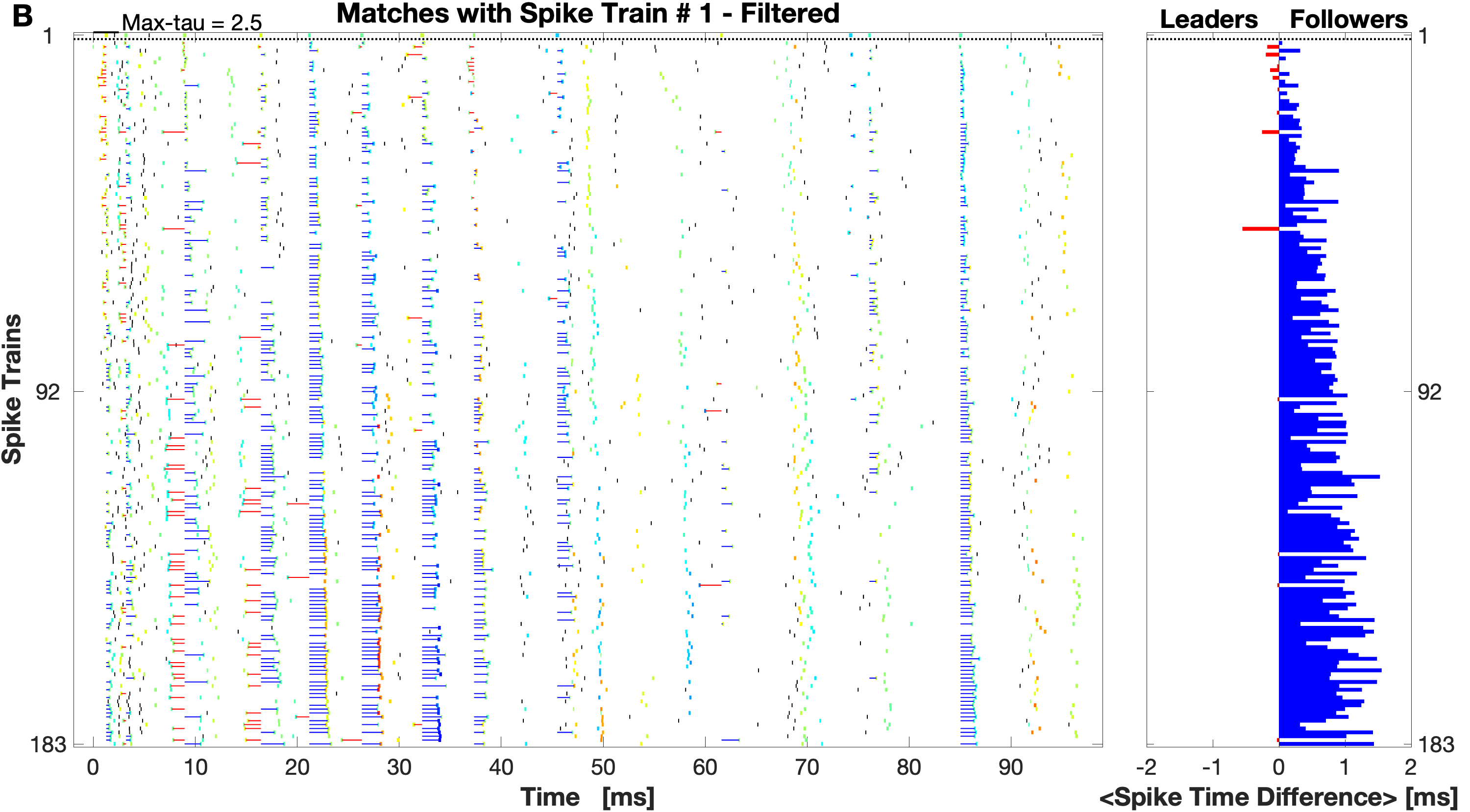}
	\end{subfigure}
	\caption{Rasterplots of the spike train set of neuron $1$ and noise $1$ of the gerbil data. In each plot there are $183$ rows corresponding to $61$ different interaural time differences (ITDs) with $3$ repetitions each. Horizontal lines identify the spike time difference of matched spikes with respect to a reference spike train (in this case the first one). Blue (red) indicates spike matches for which the order of the spikes is (not) as expected for a synfire chain, i.e., the spike of the earlier spike train comes correctly first (incorrectly last). Using the same color scheme, the histograms on the right show the average spike time differences among the matches between each spike train and the reference spike train. These are precisely the values of the first row of the Spike Time Difference Matrix. The dataset in subplot A is unfiltered and exactly as recorded, in subplot B the dataset was modified in two ways. First we added a maximum time interval of $2.5$ms between matched spikes (indicated in the upper left corner of subplot B) which reduces the number of mismatched spikes (red) considerably. We also introduced a threshold value of Spike Train Order $E = 0$ which eliminates many of the isolated background spikes from the analysis (still depicted in B but now in thin black) such that these spikes are not considered in the latency correction). As can be seen in the histograms, these two steps result in a much smoother increase of the average interval with the distance between spike trains.}
	\label{fig:Fig11_Jason_Spike_Matches}
\end{figure}

% ###############################################################################
% ######### Figure 12: Experimental Data - Example Latency Correction ###########
% ###############################################################################
%
\begin{figure*}[!thb]
    \includegraphics[width=\textwidth]{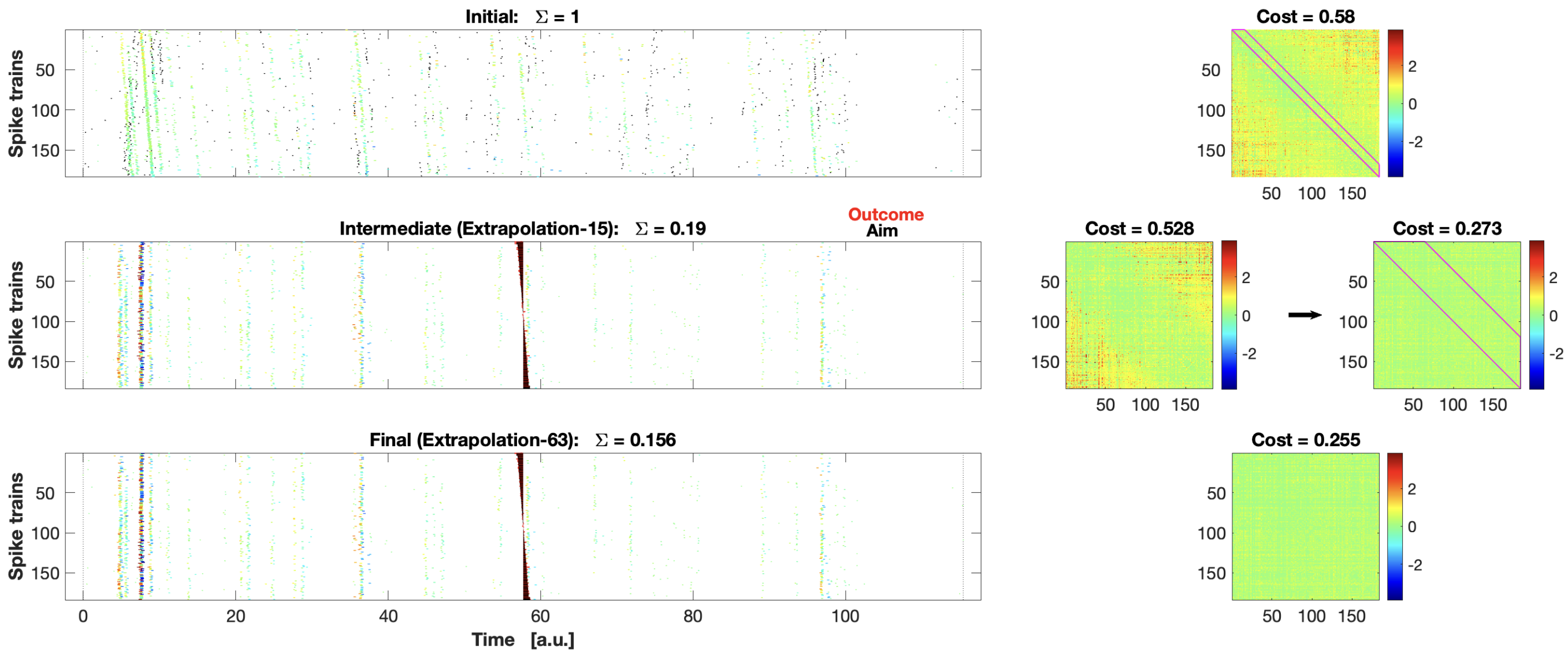}
    \caption{Latency correction for one experimental gerbil dataset (neuron 2, noise 9) which contains some global events with a rather large amount of overlap. The plot follows the iterative structure laid out in Fig. \ref{fig:Fig8-Simulated-Examples}B, only the respective areas of the STDM used in the two latency correction steps are now delineated by magenta lines. In this iterative scheme we first perform an \textit{extrapolation} direct shift with a low stop diagonal of $15$ that avoids the outer STDM regions affected by overlap. In this case there is an immediate decrease in the cost but it is only very moderate since there are still some spurious matches included in the evaluation. But this decrease gets very pronounced once the rematching is performed. This is followed by an extrapolation with stop diagonal $63$ which leads to a further improvement which can be seen in both the relative shift errors $\Sigma$ stated on the left and the cost values on the right.}
\label{fig:Fig12-Experimental-Data-Latency-Correction-Example}
\end{figure*}

% ################################################################################
% ######### Figure 13: Experimental Data - All Datasets - Overview ###############
% ################################################################################
%
\begin{figure*}[!thb]
    \includegraphics[width=\textwidth]{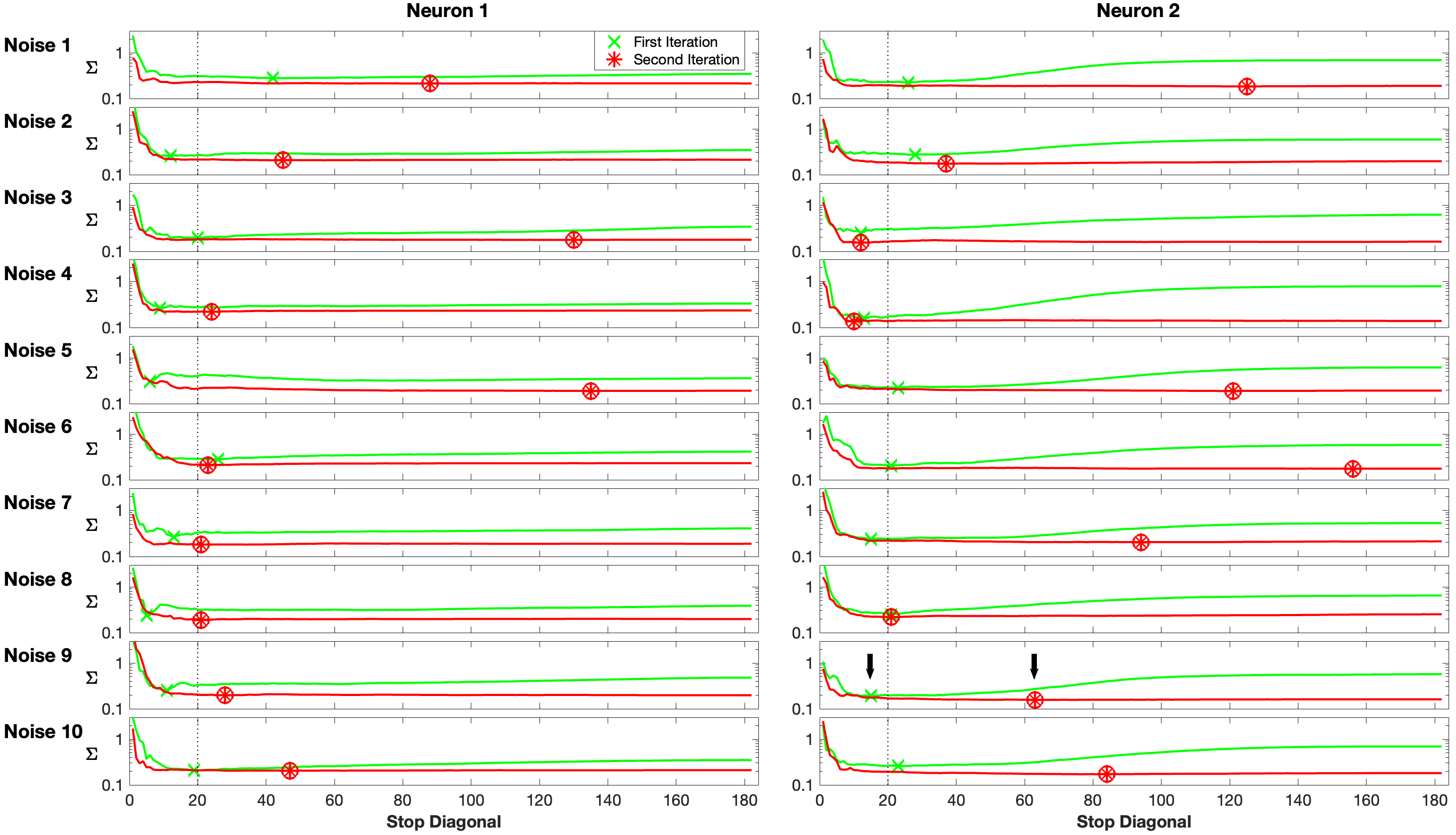}
    \caption{Relative shift error versus the stop diagonal of the Extrapolation direct shift after the first and the second iteration for the $2$ neurons and all $10$ noises. Note that we used a logarithmic y-scale to emphasize the overall decrease from the first to the second iteration. The minimum values for the two iterations are marked by green crosses and red asterisks, respectively. In all cases the overall minimum (marked by a circle) is obtained after the second iteration. The values of the example shown in Fig. \ref{fig:Fig12-Experimental-Data-Latency-Correction-Example} (neuron $2$, noise $9$) are marked by black arrows. The vertical dotted lines mark the fixed stop diagonal $\nu = 20$ that is later used for comparison.}
\label{fig:Fig13_Jason_Overview.png}
\end{figure*}

% ################################################################################
% ######### Figure 14: Experimental Data - All Datasets - Scatterplots ###########
% ################################################################################
%
\begin{figure}[h!]
    	\begin{subfigure}{0.48\textwidth}
		\includegraphics[width=\linewidth]{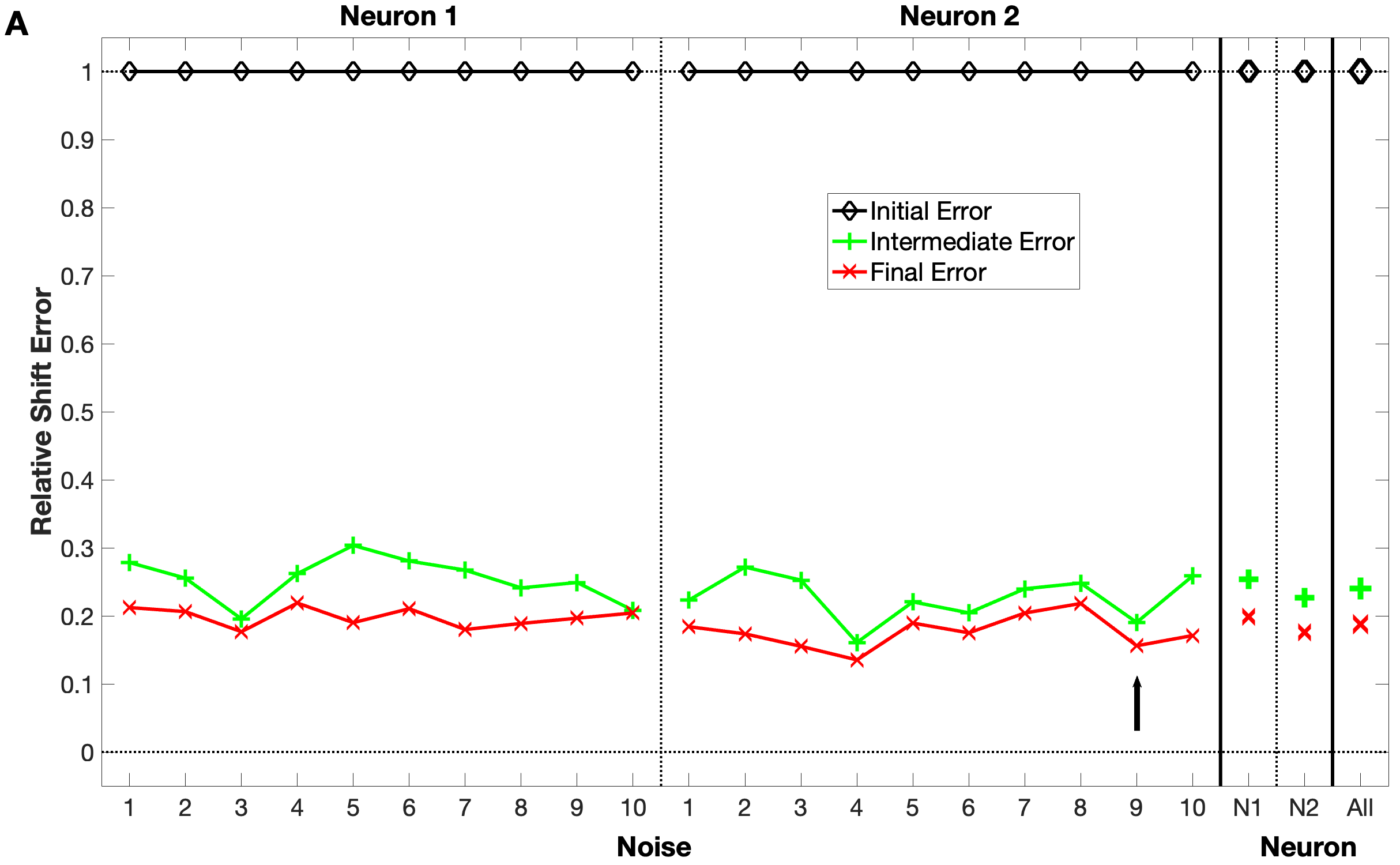}
	\end{subfigure}
	\begin{subfigure}{0.48\textwidth}
		\includegraphics[width=\linewidth]{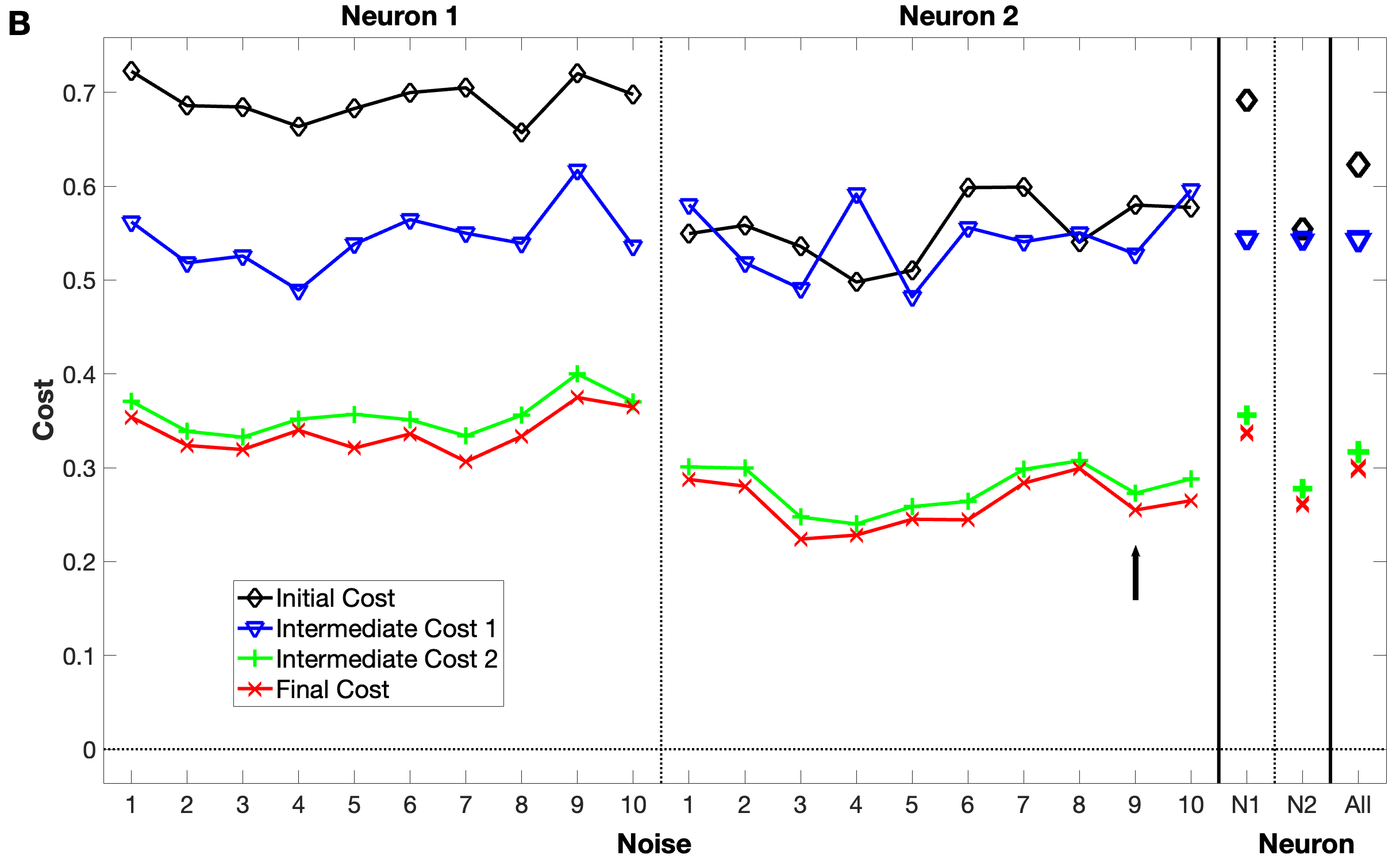}
	\end{subfigure}
	\vspace{-2mm}
	\begin{subfigure}{0.48\textwidth}	
		\includegraphics[width=\linewidth]{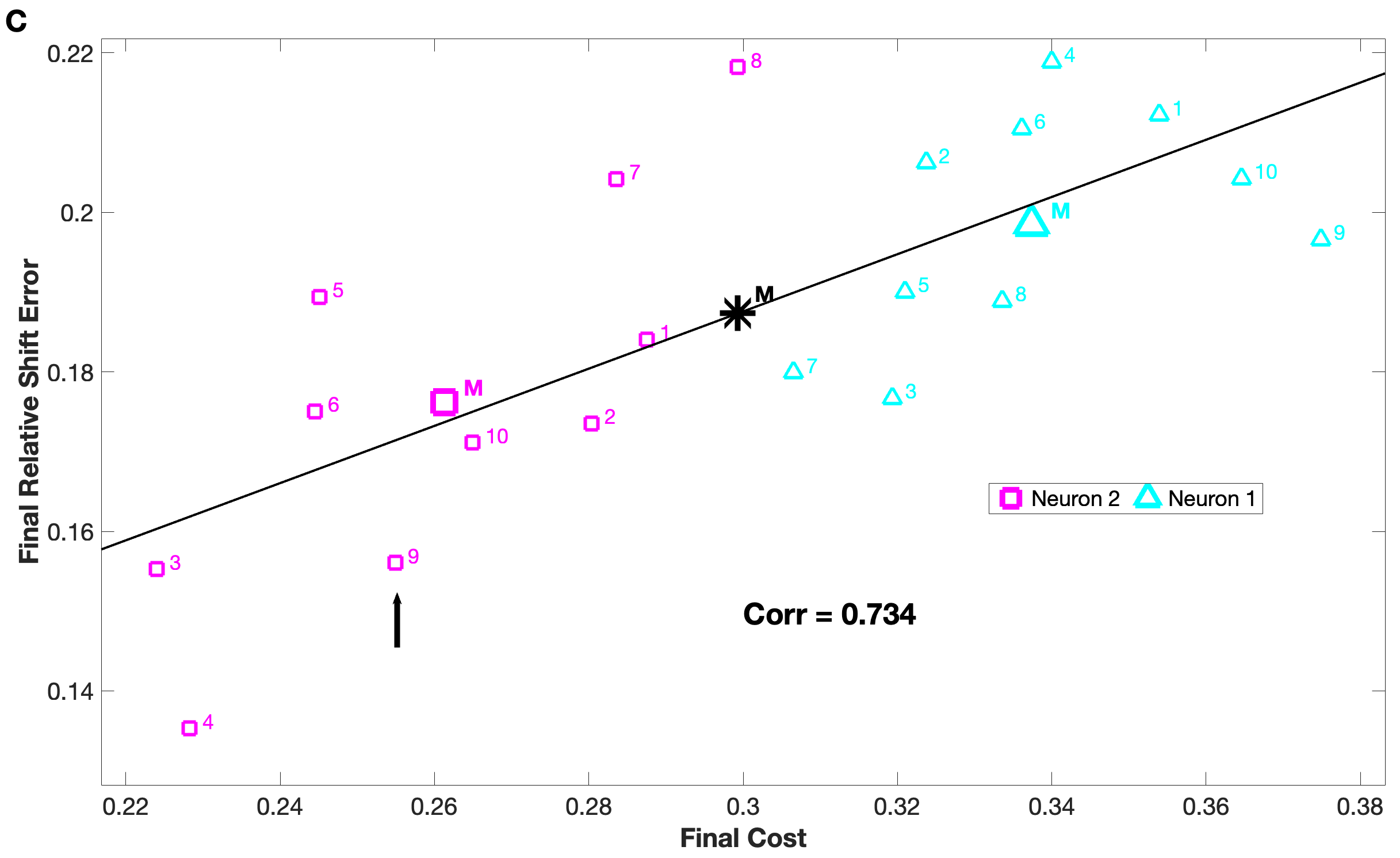}
	\end{subfigure}
    \caption{Performance of the extrapolation algorithm for all $20$ gerbil datasets. In all cases values are optimized (minimized) over all stop diagonals (see Fig. \ref{fig:Fig13_Jason_Overview.png}). In subplots A and B on the right we also show the mean values \tcr{over all $10$ noises} for neuron $1$, neuron $2$ and both neurons together, while in all three subplots the example depicted in Fig. \ref{fig:Fig12-Experimental-Data-Latency-Correction-Example} (neuron 2, noise 9) is marked by a black arrow. A. Relative shift errors. For each neuron and every noise the majority of the decrease occurs after the first iteration but in all cases the second iteration leads to a further improvement. For neuron $2$ slightly lower errors are obtained. B. Cost values. We observe the same trends as for the relative shift errors. The only differences are that there are two intermediate steps and that for neuron $2$ we now sometimes first find an increase in cost. This happens when the cost value is to a large part still determined by spurious spike matches. Rematching resolves this problem and the values after the second iteration are even lower than for neuron $1$. C. Scatterplot of final relative shift errors vs. final cost. The black line represents a linear fit, the thick markers indicate the relative mean values (black denotes overall). Overall, while there is quite some variability, neuron $2$ yields a better latency correction, which is particularly reflected in the cost for which the two ranges of values are non-overlapping. Finally, for the correlation between these two performance indicators we obtain $Corr = 0.734$.}
	\label{fig:Fig14_Jason_Statistics}
\end{figure}

After having validated the methods developed in Section \ref{s:Methods2} on simulated data, we now apply them to real neurophysiological data, namely single-unit recordings from two medial superior olive (MSO) neurons of an anaesthetised gerbil during presentations of ten different noisy auditory stimuli \cite{Beiderbeck22} (for a description of the data please refer to Section \ref{s:Data} and \ref{App1-Experimental-Setup}). For every neuron and each noise ($2 \times 10 = 20$ datasets) we have $3$ repetitions of $61$ different angles of the noise source (thus varying the interaural time difference, ITD) yielding a total of $183$ spike trains. All of these datasets contain many global events with varying degrees of overlap and, equally important, different but quite large levels of noise.

% XXXXX Paragraph on validation, ground truth, plus strategy from here on XXXXX

In response to this noisiness, we carry out two preprocessing steps that both aim at denoising the data. Their combined effect is illustrated in Fig. \ref{fig:Fig11_Jason_Spike_Matches}. The first problem that can throw off the adaptive coincidence detection and lead to unintended consequences is the presence of incomplete global events surrounded by sparse spiking. Under these circumstances spikes from the global event often get (erronously) matched with spikes very far away from that event. The easiest way to prevent this is to set a maximum distance between two spikes from the same global event. After estimating the overall interval from the first to the last spike for the slowest events we chose a \tcr{higher} value \tcr{of} $\tau_{max} = 2.5$ms which eliminates the more excessive mismatches, as can be seen when comparing the number of mismatched spikes (red) and the length of the intervals between them in subplots A and B in Fig. \ref{fig:Fig11_Jason_Spike_Matches}. In the second iteration we choose only half that value (i.e., $\tau_{max} = 1.25$ms), since after the first iteration spikes are already quite well aligned and the duration of even the slowest global event is significantly reduced.

The second preprocessing step consists of a filtering of the spike trains by means of the Spike Train Order profile $E$ (Eq. \ref{Eq:Spike-Train-Order-E-Spike-1}). A similar filtering of the spike trains (but based on the SPIKE-synchronization C) was already performed in \citep{Kreuz17}. The aim here is to focus the analysis on the spikes within the global events and ignore isolated spikes that do not really contribute to the synfire chain within the datasets. In order to not alter the data too much we set a rather conservative filter of $E = 0$, which means that we only eliminate spikes that participate in inverse synfire chains. Typically this will be sparse spikes in between two global events that also happen to be in the wrong order to each other. After this double filtering the average interval increases much more smoothly with the distance between spike trains and in particular there are much less average intervals in the wrong direction (see histograms in Fig. \ref{fig:Fig11_Jason_Spike_Matches}).

In Fig. \ref{fig:Fig12-Experimental-Data-Latency-Correction-Example}, we show iterative latency correction, as described in Section \ref{ss:Methods_2-2_New-Algorithms}, exemplarily applied to an experimental dataset (neuron 2, noise 9) that first underwent this kind of double filtering. Since simulated annealing is not feasible for datasets with such an elevated number of spike trains, we restrict ourselves to the much faster extrapolation direct shift \footnote{\tcr{The computational cost of both algorithms depends on the number of spike trains and spikes in the data. On a machine with an M1 processor, 8GB RAM and Matlab 2024b we applied both simulated annealing and the extrapolation direct shift to an experimental dataset with $183$ spike trains and an average number of 13 spikes each. Even using Matlab EXecutables (MEX) compilation to speed up simulated annealing it took $861.8$ seconds, as compared to only $9.7$ seconds for direct shift. Therefore, we observe an approximately $88$-fold speed increase for the newly proposed extrapolation method. Since calculations are based on the $N(N-1)/2$ values of the spike time difference matrix, both methods scale quadratically with the number of spike trains, e.g., $o(N^2)$. When we extrapolate towards the asymptotic ratio of their running times for very large numbers of spike trains, we obtain an ultimate speed gain factor of $130.7$.}} which, in any case, performs very similar (compare Fig. \ref{fig:Fig10-Simulations-Overall-Comparison}). As we have done in the simulated example shown in Fig. \ref{fig:Fig8-Simulated-Examples}B, we start our iterative scheme with a rather low stop diagonal (here not $1$ but $15$) which helps minimizing the effect of the spurious spike matches visible in the off-diagonal corners of the STDM. This first iteration reduces the relative shift error considerably from $1$ to $0.19$ and the cost value very modestly from $0.58$ to $0.528$) which is caused by the inclusion of still spurious spike matches. However, rematching the now much better aligned spikes brings the cost down to $0.273$, and we can still improve upon this by applying a second round of extrapolation direct shift, now with a much larger stop diagonal of $\nu = 63$. This second iteration decreases ultimately both the shift error (to $0.156$) and the cost value (to $0.255$).

Note that in the example of Fig. \ref{fig:Fig12-Experimental-Data-Latency-Correction-Example} for both iterations we optimized (i.e., minimized) the relative shift error over all possible stop diagonals. In Fig. \ref{fig:Fig13_Jason_Overview.png} we show the whole range of stop diagonals not only for this example but for all $20$ datasets from both neurons and across all the different noise stimuli. In all these cases the traces of the two iterations exhibit a very similar course. The relative shift errors for low stop diagonals are quite high but then the error tends to descend towards a rather broad minimal plateau and a generally slow but sometimes, in particular for neuron $2$, more pronounced increase for high stop diagonals. Apart from very few exceptions, the minimum for the first iteration is obtained for lower stop diagonals than for the second iteration and in absolutely all cases it is the second iteration that yields the overall minimum.

These results for the relative shift errors are summarized in Fig. \ref{fig:Fig14_Jason_Statistics}A. For all $20$ dataset the decrease from the initial relative shift error (always $1$ by definition, as already explained in Section \ref{ss:Methods_2-3_Relative-Shift-Error}) is very pronounced for the first iteration and less prominent for the second iteration but in all cases the final value is the absolute minimum. The same holds true for the cost values depicted in Fig. \ref{fig:Fig14_Jason_Statistics}B. Here, in addition, for neuron $2$ the first intermediate step for noises $1$, $4$, $8$, and $10$ leads to an increase in cost. This is similar to the example shown in Fig. \ref{fig:Fig8-Simulated-Examples}B and happens when the cost value is still mostly based on spurious spike matches. The problem is resolved by the intermediate rematching step and in fact the final cost values are even better (i.e., lower) than for the first neuron. For both indicators neuron $2$ yields better latency correction performance values, a fact that can be appreciated even better in Fig. \ref{fig:Fig14_Jason_Statistics}C where we plot one against the other. For the cost the separation between the two neurons is indeed perfect since even the highest value of neuron $2$ is smaller than the lowest value of neuron $1$. The same scatterplot can also be used to look at the correlation between the two indicators of latency correction performance and here we find the value of $Corr = 0.734$, quite high in particular when considering the substantial noisiness of the data.

In this study we analyze experimental data with a known ground truth which allows us to employ the relative shift error as a performance measure for the latency correction and use it to optimize over the free parameter, the stop diagonal. An important advantage of our extrapolation direct shift method is its cumulative nature (the outer diagonals are (re-)constructed from sums of the inner diagonals) which leads to a very large degree of smoothness in the dependence on the stop diagonal (as can be seen in Fig. \ref{fig:Fig13_Jason_Overview.png}). This is in contrast to another method that we tried (termed intrapolation) which was much noisier and fluctuating. While it occasionally achieved lower minimum values, it did so in a very unpredictable manner such that its practical usefulness for datasets with unknown ground truth was very limited (which is why its definition and its results are not included here). On the other hand, due to the smoothness of the extrapolation algorithm, the importance of the exact value of the stop diagonal is diminished considerably which is exactly what makes this method applicable to datasets without known ground truth for which we can not use the relative shift error but instead have to rely on the reduced matrix cost value as measure of performance.

A useful guideline is to run the first iteration with the stop diagonal set at a rather low value (of about $10 - 20 \%$ of the number of spike trains) and then, once the effect of any initial overlap is eliminated (or at least substantially reduced), use the second iteration to optimize the cost value over all stop diagonals. In fact, when we did this for the gerbil data, with a fixed initial stop diagonal of $\nu = 20$ we got a performance reduction - averaged over all $20$ spike train sets that beforehand were individually optimized - of only $5.87 \%$ ($0.198$ versus $0.187$). This demonstrates the robustness of the extrapolation algorithm against uncertainties in the selection of the stop diagonal parameter. 

% XXXXX As a final step of validation, we also analyzed a third completely new dataset XXXXX

% #############################################################################
% #################### Section: Conclusions ###################################
% #############################################################################
%
\section{Conclusions} \label{s:Conclusions}

% Summary
In studying the nature of global events in neuronal datasets the latencies presented by the spatio-temporal propagation of the signal can constitute a hindrance for the proper estimation of synchronization in the data. \tcr{Of course, such delays usually have a physiological meaning, but eliminating them allows to analyze whether there is a loss of information in the transfer of activity between neurons or brain areas, or in the alternative neuronal coding scenario, to investigate how far the response to repeated presentation of a stimulus depends on the onset latency.} 

Recently latency correction algorithms have been proposed to resolve this problem \citep{Kreuz22}. However, when spikes from different global events exhibit overlap, i.e., when an event starts before the previous one has ended, and even before (as soon as spurious relations between spikes from different events are inferred), new problems arise. These are caused by the underlying coincidence detection algorithm \cite{Kreuz17} which uses temporal vicinity as the criterion to establish spike matching. In this study we overcome these limitations by deriving a new theory of event overlap and designing novel algorithms of latency correction capable of dealing with overlapping events.

In the Methods part we first build on the SPIKE-Order framework \cite{Kreuz17} and quantify the amount of overlap in the dataset in terms of the stop diagonal of the spike time difference matrix (STDM). After considering the invariance under homogeneous time translations we propose the relative shift error as an advanced performance measure of such algorithms for the case in which a ground truth is known. Next, we combine two latency correction algorithms initially proposed in \citep{Kreuz22}, the very fast extrapolation direct shift and the much slower simulated annealing. Both algorithms make use of the reduced STDM only, i.e., without the outer off-diagonal parts that are most affected by spurious matches due to event overlap. Finally, the algorithms are embedded in a newly developed iterative scheme that uses in each step as much of the latency information as possible. In all of these cases the stop diagonal acts as the crucial parameter.

%Results
In the Results we then first use simulated data with a known ground truth to validate and to compare the new latency correction algorithms with previous approaches regarding sensitivity to event overlap and robustness to noise. We notice that there is an initial tradeoff between these two quantities since no method works best under all circumstances. However, we also find that this tradeoff can be overcome by means of an iterative scheme which works best when the first iteration is restricted to low step diagonals so that the overlap is taken care of and the second iteration is more expansive and includes more information from the STDM which optimizes also with regard to noise. We also discover that extrapolation direct shift, although much faster, works basically as good as simulated annealing. Therefore, in the much larger experimental datasets (single-unit recordings from two medial superior olive neurons of a gerbil) where simulated annealing is not feasible, we restrict ourselves to the computationally much less demanding extrapolation method.

\tcr{In these data the overlap arises due to the inter-event intervals of global events within traces (determined by the phasic nature of inputs to the MSO neurons) and the onset latency of the aforementioned events across traces (due to the implementation of \emph{interaural time differences} (ITDs), i.e., how the stimulus was accelerated or delayed at either ear). These ITDs, small deviations in the arrival times of sounds between the two ears of the gerbils, enable the animals to spatially localize the noise source (see Section \ref{s:Data}).} This overlap can now be resolved and we are able to disentangle successive global events. Again, the iterative scheme helps to push the remaining shift error as low as possible. Finally, the new extrapolation method also proves to be very robust against changes in the only algorithm parameter, the stop diagonal, which is very important when applying these methods to real datasets without any ground truth.

% Impact on data
\tcb{For the gerbil data this technique offers a novel way to quantify the impact of monaural filtering on the temporal dynamics of binaural coincidence detection in the Medial Superior Olive nucleus. Acoustic stimuli arriving at each ear are filtered at the cochleae, leading to bandpass-filtered monaural inputs whose envelopes, i.e., their fluctuating energy, help determine the timing and probability of binaural responses \cite{agapiou2008low, brughera2022sensitivity, dietz2014emphasis}. Previous MSO studies have required presentation of less-ethological, tonal stimuli (binaural beats \cite{franken2015vivo, van2013directional} or the “Zwuis” tone complex \cite{mackenbach2023somatic, plauvska2016predicting}) to identify the contribution of monaural filtering to binaural processing. However, the associated analysis techniques remove the temporal firing patterns from the MSO data (often even ignoring the onset of the response \cite{franken2015vivo}), thus they disregard how the envelope, especially its rising energy \cite{dietz2014emphasis}, of cochlea-filtered signals can impact MSO responses and therefore binaural processing. By allowing us to identify global events associated with a dominant ear from extracellular recordings of ethological stimuli, this new technique offers the possibility to perform reverse correlation on the suprathreshold events and predict the underlying cochlea-filtered acoustic stimuli.}

% Universal applicability: Different spatio-temporal scales and kinds of data

Regarding the general applicability, algorithms of latency correction work with arbitrary spatial and temporal scales. Here we are operating within a very microscopic framework but examples of more large-scale datasets that can be represented by rasterplots include spatio-temporal propagation of activity in the electroencephalogram (EEG) of epilepsy patients (where the discrete events are obtained from local maxima and minima of the oscillatory EEG signals) \citep{Kreuz13} and highly organized patterns of spontaneous fluctuations as measured via peak detection in resting-state functional magnetic resonance imaging (fMRI) \citep{gu2021brain}. Furthermore, while the majority \tcr{of} our previous studies on spike train analysis have been carried out within a neuroscientific context, we would like to stress that these latency correction algorithms are universal and can be applied to discrete datasets from many scientific fields. These include fields as diverse as for example climatology (where the measured scalar could be the change and propagation of temperature, rainfall etc.) \citep{Kreuz17, sun2018patterns, Conticelli20}, network science \citep{mwaffo2018detecting, Bardin19}, social communication \citep{Varni10}, mobile communication \citep{wang2022identifying} and policy diffusion \citep{grabow2016detecting}.

% Limitations & Outlook
For the future there remain a few directions that can build on the progress made in this article. One avenue is to improve the suitability of the algorithm for very sparse spike trains. This even includes datasets which contain pairs of spike trains without any matches (SPIKE-synchronization $C^{(n,m)} = 0$), the effects of which we have seen in subplots C, E, and G of Fig. \ref{fig:Fig9-Simulated-3D-Plots}. The important point here is that these matchless spike train pairs should not necessarily be interpreted as an incentive not to shift at all. Rather it is better to exclude the respective elements of the spike time difference matrix from the analysis. This amounts to a different kind of filtering and here some work remains still to be done to find the right balance between minimum statistical requirements and exclusion of insufficient spike trains from the analysis. One promising approach seems to be weighing of different entries of the STDM.

% Current Problems
In \citep{Kreuz22} we could show that ‘‘standard" latency correction works best for sparse data with well-defined global events (as manifested by high values of SPIKE-synchronization $C$) and a consistent order within these events (corresponding to elevated values of the Synfire Indicator $F$). For experimental datasets with considerable overlap, such as the ones analyzed here, both of these quantities are tainted by spurious matches, and a more careful analysis of a much broader database will be required to find out under which circumstances the latency correction of such datasets yields the best results and where the limits are. What is already clear is that when the overlap reaches the theoretical limit $R = (N - 1) / 2$ defined in Section \ref{ss:Methods_2-1_Overlap-Theory} all latency correction based on event matching becomes futile.

Regarding the estimation of synchrony after the latency correction, instead of looking just at the cost function or SPIKE-synchronization as measures of spike matching, one could use more comprehensive measures of spike train similarity, for example spike train distances \cite{Kreuz20} such as the time-scale dependent Victor-Purpura \cite{Victor96} and van Rossum \cite{VanRossum01} distances or the time-scale independent ISI- \cite{Kreuz07c} and SPIKE-distances \cite{Kreuz13}.

% Source Codes (Rephrased from 2022 paper)
Optimized implementations of the algorithm will be made available in three freely available software packages. These are the Matlab graphical user interface SPIKY\footnote[1]{http://www.thomaskreuz.org/source-codes/SPIKY} \cite{Kreuz15}, the Python library PySpike\footnote[2]{http://mariomulansky.github.io/PySpike} \cite{Mulansky16} and Matlab command line library cSPIKE\footnote[3]{http://www.thomaskreuz.org/source-codes/cSPIKE}. The new latency correction algorithms will complement the existing content of these software packages such as the three symmetric measures of spike train synchrony, ISI-distance \cite{Kreuz07c, Kreuz09}, SPIKE-distance \cite{Kreuz11, Kreuz13} and SPIKE-synchronization \cite{Kreuz15} (see \cite{Satuvuori17} for adaptive generalizations), the directional SPIKE-Order \cite{Kreuz17} as well as algorithms that find within a larger neuronal population the most discriminative subpopulation \cite{Satuvuori18b}.

%\section*{Acknowledgement}
%
%We thank XXXXX for useful discussions and XXXXX for a careful reading of the manuscript. This project has received funding from XXXXX.
%
% #############################################################################
% ################################ Appendix ###################################
% #############################################################################
%
\begin{appendix}

\section*{Appendix \label{Appendix}}

Here we present the technical details of the data recordings (\ref{App1-Experimental-Setup}), the mathematical definitions of SPIKE-synchronization (\ref{App2-SPIKE-synchronization}) and SPIKE-Order (\ref{App3-SPIKE-Order-Synfire-Indicator}) and an in-depth look at the newly defined Relative Shift Error (\ref{App4-Relative-shift-error}):

\tcb{\section{Experimental setup and data collection \label{App1-Experimental-Setup}}}

\tcb{All animal experiments were approved in accordance with the stipulations of the German animal welfare law (55.2-1-54-2532-53-2015). In vivo, single-unit recordings were performed in $3-7$ month old Mongolian Gerbils (Meriones Unguiculatus) of either sex who were anaesthetized with a Ketamine/Xylazine mix (Ketamine $20\%$, MEDISTAR, GmbH; Xylazine, $2\%$ Bayer AG diluted in $0.9\%$ NaCl solution).} 

\tcb{The gerbils were laid on thermostatically controlled heat pad (Fine Science Tools GmbH) in a sound-attenuated chamber and head-fixed within a custom-made stereotaxic frame \citep{Beiderbeck18, Beiderbeck22, muller2023temporal}. Two craniotomies were performed: the first one between bregma and lambda to place the reference electrode and the second behind the sinus transversus lateral to the midline which, in conjunction with a durotomy, allowed access to the brainstem for recordings.}

\tcb{Extracellular recordings of action potentials (APs) from Medial Superior Olive neurons were performed using glass electrodes (Sigma-Aldrich) filled with $5$ units/$\mu$l horseradish peroxidase (HRP) diluted in $10\%$ NaCl solution, leading to a tip resistance of $~5-20M\Omega$. Their descent into the brainstem at a $20$deg angle was controlled using a motorized micromanipulator (Mitutoyo) and piezo-drive (Inchworm controller $8200$, EXFO Burleight Products Group).}

\tcb{Calibrated earphones (ER-4 microPro, Etymotic Research) were placed on the ear canals of the gerbils with their heads stereotactically aligned relative to lambda. The stimuli were generated in MATLAB (MathWorks) at a sampling rate of $192$kHz and were played via a sound card interface (Fireface UFX, RME-Audio) using the AudioSpike software package (HörTech). To each MSO neuron $10$ distinct, frozen white noise tokens that were binaurally correlated were presented. Their duration was $100$ms with $5$ms cos-ramps at the onset and offset. The tokens were played at $30$dB above threshold which led to a range from $40-65$ dB SPL. Each noise token was presented at $61$ ITDs from $-1.4$ms (noise leading at the ear ipsilateral to the recorded unit) to +$1.4$ms (contralateral ear) in $0.05$ms increments (plus four more values at $\pm1.43$ and $\pm1.79$). Per noise token and ITD $3$ repetitions were performed in a pseudo-random fashion.}
% $61$ ITDs, ranging from $±1.79$ms, with a $0.05$ms step-size used between $±1.4$ms.

\tcb{Extracellular voltage (and voltage changes associated with APs) were measured with a pre-amplifier (Electro $705$, World Precision Instruments), filtered (Hum Bug Noise Eliminator, Quest Scientific Instruments), converted and then delivered to the computer via the sound card (Fireface UFX, RME-Audio). APs were analyzed online using the AudioSpike (HörTech). Single-units were tested by visual inspection and online sorting. A signal-to-noise ratio of the spike waveform of $> 5$ was required for recorded neurons to be included for further analysis. To seek responsive neurons $200$ms white noise bursts were used,  while MSO neurons were identified during recordings by their characteristic ‘EE’ response, i.e., stimulation of the ipsilateral and contralateral ear evoking neuronal spiking. This was later confirmed by iontophoretic release of the HRP followed by subsequent labelling of the recordings site with a $3, 3'$-diaminbenzidine (DAB) substrate kit for peroxidase (Vector Laboratories). MSO data were analysed using custom-made programs in MATLAB (MathWorks). Considering the stimulus onset at $0$ms, all AP timings were advanced by $14$ms to account for delays associated with presentation of the stimuli and the corresponding cochlear/multisynaptic delays in the MSO pathway.}

\section{SPIKE-synchronization \label{App2-SPIKE-synchronization}} 

To arrive at an overall measure of spike matching for any given spike train set, we start from the adaptive coincidence criterion \cite{QuianQuiroga02b} of Eq. \ref{Eq:Coincidence-MaxDist} and the coincidence indicator defined in Eq. \ref{Eq:Coincidence-Indicator} (see Section \ref{ss:Methods_1-1_Coincidence-Detection}). First we calculate a multivariate normalized coincidence counter for each spike of every spike train
\begin{equation} \label{Eq:SPIKE-sync-Multi-Counter}
	C_i^{(n)}=\frac{1}{N-1} \sum_{m \neq n} C_i^{(n,m)}
\end{equation}
by averaging over all $N-1$ bivariate coincidence indicators involving the spike train $n$.

Subsequently, in order to obtain a single multivariate SPIKE-synchronization profile we pool the coincidence counters of the whole spike train set:
\begin{equation} \label{Eq:SPIKE-sync-Multi-Profile}
    	\{C(t_k)\} = \bigcup_n \{C_{i(k)}^{(n(k))} \},
\end{equation}
where we map the spike train indices $n$ and the spike indices $i$ into a global spike index $k$ denoted by the mapping $i(k)$ and $n(k)$. 

With $M$ denoting the total number of spikes in the pooled spike train, the average of this profile
\begin{equation} \label{Eq:SPIKE-synchronization-C}
	C = \begin{cases}
		\frac{1}{M} \sum_{k=1}^M C(t_k) & {\rm if} \  M > 0 \cr
		\ \ \ \ \ \ \ \ \ 1 & {\rm otherwise}
	\end{cases}
\end{equation}
\noindent yields the SPIKE-synchronization value $C$, the overall fraction of coincidences \cite{Kreuz15}. It reaches the value $1$ if and only if each spike in every spike train has one matching spike in all the other spike trains (or if there are no spikes at all - since common silence is also perfect synchrony), and it attains the value $0$ if and only if the spike trains do not contain any coincidences.

\section{SPIKE-Order, Spike Train Order and the Synfire Indicator \label{App3-SPIKE-Order-Synfire-Indicator}} 

SPIKE-synchronization is invariant to which of the two spikes within a coincidence pair is leading and which is following. To take the temporal order of the spikes into account we developed the SPIKE-Order approach \cite{Kreuz17} which allows to sort the spike trains from leader to follower and to evaluate the consistency of the preferred order via the Synfire Indicator.

We first define the bivariate anti-symmetric SPIKE-Order 
\begin{eqnarray} \label{Eq:SPIKE-Order-Spike-D}
	D_i^{(n,m)} & = & C_i^{(n,m)} \cdot \sign (t_{j'}^{(m)} - t_i^{(n)}) \nonumber \\
	D_{j'}^{(m,n)} & = & C_{j'}^{(m,n)} \cdot \sign (t_i^{(n)} - t_{j'}^{(m)}) = - D_i^{(n,m)},
\end{eqnarray}
which assigns to each spike $i$ either a $+1$ or a $-1$ depending on whether the respective spike is leading or following the coincident spike $j'$ in the other spike train (cf. Eq. \ref{Eq:Matching-Spike}). SPIKE-Order distinguishes leading and following spikes, and is thus used to colorcode the individual spikes on a leader-to-follower scale (see, e.g., Fig. \ref{Fig1:Overview-Rasterplots}). The profile is invariant under exchange of spike trains, i.e., it looks the same for all events no matter what the order of the firing is. Moreover, summing over all profile values (which is equivalent to summing over all coincidences) necessarily leads to an average value of $0$, since for every leading spike $(+1)$ there has to be a following spike $(-1)$.

Spike Train Order $E$ is similar to SPIKE-Order $D$ but with two important differences: Both spikes are assigned the same value and this value depends on the order of the spike trains:
\begin{equation}  \label{Eq:Spike-Train-Order-E-Spike-1}
 E_i^{(n,m)} = C_i^{(n,m)} \cdot
 				\begin{cases}
 					\sign (t_{j'}^{(m)} - t_i^{(n)})\quad\text{if}\quad n<m\\
                		\sign (t_i^{(n)} - t_{j'}^{(m)})\quad\text{if}\quad n>m
              	\end{cases}
\end{equation}
and
\begin{equation} \label{Eq:Spike-Train-Order-E-Spike-2}
	E_{j'}^{(m,n)} = E_i^{(n,m)}.
\end{equation}

The spike trains are sorted by means of the cumulative and anti-symmetric Spike Train Order matrix
\begin{equation} \label{Eq:Spike-Train-Order-Matrix}
    E^{(n,m)} = \sum_i E_i^{(n,m)}
\end{equation}
which quantifies the temporal relationship between spike trains $n$ and $m$. 
If $E^{(n,m)}>0$ spike train $n$ is leading $m$, while $E^{(n,m)}<0$ means $m$ is the leading spike train. For a Spike Train Order in line with the synfire property (i.e., exhibiting consistent repetitions of the same global propagation pattern), we thus expect $E^{(n,m)} > 0$ for all $n<m$. Therefore, the overall Spike Train Order can be constructed as
\begin{equation} \label{Eq:Spike-Train-Order}
 	E_< = \sum_{n<m} E^{(n,m)},
\end{equation}
i.e.\ the sum over the upper right tridiagonal part of the matrix $E^{(n,m)}$.

Finally, normalizing this cumulative quantity by the total number of possible coincidences yields the Synfire Indicator:
\begin{equation} \label{Eq:Synfire-Indicator-F}
	F = \frac{2 E_<}{(N-1) M}. 
\end{equation}
This measure quantifies to what degree coinciding spike pairs with correct order prevail over coinciding spike pairs with incorrect order, or, in other words, to what extent the spike trains in their current order resemble a consistent synfire pattern.
Accordingly, maximizing the Synfire Indicator $F_\varphi$ as a function of the spike train order $\varphi(n)$ finds the sorting of the spike trains from leader to follower such that the sorted set $\varphi_s$ comes as close as possible to a perfect synfire pattern \cite{Kreuz17}:
\begin{equation} \label{Eq:Sorted-Order}
	\varphi_s: F_{\varphi_s} = \max_\varphi \{F_\varphi\} = F_s.
\end{equation}

Whereas the Synfire Indicator $F_\varphi$ for any spike train order $\varphi$ is normalized between $-1$ and $1$, the optimized Synfire Indicator $F_s$ can only attain values between $0$ and $1$. But from Eq. \ref{Eq:Spike-Train-Order-E-Spike-1} it follows that since the order is only evaluated among those spikes that match each other, the actual upper bound for any given dataset is the value of SPIKE-synchronization $C$ (Eq. \ref{Eq:SPIKE-synchronization-C}). A perfect synfire pattern results in $F_s=1$, while sufficiently long Poisson spike trains without any synfire structure yield $F_s \gtrsim 0$.

\section{Relative shift error calculations and metrics \label{App4-Relative-shift-error}}

As introduced in Methods Section \ref{ss:Methods_2-3_Relative-Shift-Error}, the relative shift error $\Sigma$ enables us to define a meaningful error function which respects all the time translation symmetries inherent to the latency correction algorithms we designed. To start, it is  fundamental to specify the metric utilized to calculate the norm of the shift vectors. Each shift should in fact have a norm invariant under homogeneous time translation, as the shift should only represent the relative distance between spikes and not their group difference from the origin. Using simple metrics such as Euclidian or Taxicab will not return unambiguous results as they do not respect invariance under homogeneous time translation.

To express this invariance any single shift should more aptly be represented by a line parallel to the identity vector in N-dimensional space such that
\begin{align}
	\vec{s}  \; \rightarrow & \; \vec{s} - \lambda \vec{\mathds{1}} \quad \forall \; \lambda \; \in \; \mathds{R}  \\
	\vec{\mathds{1}} &= \begin{bmatrix}
		1 \\
		1 \\
		\vdots \\
		1
	\end{bmatrix}.
\end{align}
An intuitive metric for this invariant \emph{shift} object could consist in minimizing the Euclidian metric along the line by projecting each shift onto the $N-1$-dimensional subspace orthogonal to the identity vector (as illustrated in Fig. \ref{Fig7:Relative-Shift-Error}). Thus every shift will be such that
\begin{equation}
	(\vec{s} - \lambda_{min}\vec{\mathds{1}}) \cdot \vec{\mathds{1}} = 0 ,
\end{equation}
which implies
\begin{equation}
	\lambda_{min} = \frac{1}{N} \sum_{i=1}^{N} \vec{s}_i.
\end{equation}
Hence the absolute time translation that projects the shifts onto the $N-1$-dimensional subspace orthogonal to $\vec{\mathds{1}}$ and thus minimizes the Euclidian norm is the average of the shift itself. A similar argument holds for the taxicab norm. Here the absolute minimum is obtained for the median value of the shift vector ($\lambda_{min} = median(\vec{s})$). In fact
\begin{equation}
	\lvert \lvert \vec{s} - \lambda \vec{\mathds{1}} \rvert \rvert_{taxicab} = \sum_{i=1}^{N} \abs{\vec{s}_i - \lambda \cdot 1 }
\end{equation}
and by differentiating $\lambda$ and setting the derivative to $0$ in search of the minimum we find
\begin{equation}
	\sum_{i=1}^{N} \frac{\vec{s}_i - \lambda}{\abs{\vec{s}_i - \lambda}} = 0,
\end{equation}
which is satisfied for $\lambda_{min}$ equal to the median of $\vec{s}$. These two statements relate to the well-known result that the mean and median are the estimators of $L_1 \; and \; L_2$, respectively \citep{cormen1990introduction}.

In general for any chosen shift $\vec{s}$ and metric there will be such a shift $\vec{s^{\lambda}}$ obtained through the homogeneous translation which minimizes the vector's norm
\begin{equation}
	\vec{s^{\lambda}} = \vec{s} - \lambda_{min} \vec{\mathds{1}}
\end{equation}
with 
\begin{equation}
	\lambda_{min} = min_{\lambda}( \lvert \lvert \vec{s} - \lambda \vec{\mathds{1}} \lvert \lvert).
\end{equation}
In this way we obtain two norms that respect the homogeneous time invariant nature of the shifts, by first translating the vector by the $\lambda_{min}$ value respective to the norm utilized (Euclidian or Taxicab) and then calculating the resulting norm using the chosen metric. As already mentioned in Section \ref{ss:Methods_2-3_Relative-Shift-Error}, in this paper we use the taxicab metric.

After having extrapolated a complete spike time difference matrix (see Section \ref{ss:Methods_2-2_New-Algorithms}) a shift can be obtained by averaging over the columns/rows
\begin{equation}
	\vec{s} =  \sum_{n} \frac{\delta^{(n, m)}}{N} = \sum_{\alpha} \frac{\vec{s^{\alpha}}}{N}
\end{equation}
where $\vec{s^{\alpha}}$ is the shift vector in the reference frame of the $\alpha-th$ spike train, meaning where all spikes are shifted according to their distance from the $\alpha$ spike train and the $\alpha-th$ element of the shift is by definition 0: $\vec{s^{\alpha}}_\alpha = 0$ (corresponding to the diagonal elements of the STDM).

It is possible to obtain the shift in the $\alpha$ spike train reference frame from the shift in the absolute reference frame which minimizes the Euclidian Distance $\vec{s^{\lambda}}$ by simply subtracting its $\alpha-th$ element
\begin{equation}
	\vec{s^{\alpha}} = \vec{s^{\lambda}} - s^{\lambda}_{\alpha}\vec{\mathds{1}}
\end{equation}
as this implies that the $\alpha-th$ element of the vector is $0$. Summing over the different alphas we obtain
\begin{equation}
	\sum_{\alpha}\vec{s^{\alpha}} = \sum_{\alpha}  ( \vec{s^{\lambda}} - \vec{s^{\lambda}}_{\alpha}\vec{\mathds{1}} ).
\end{equation}
Since the last term of this equation is $0$, the invariant shift $\vec{s^{\lambda}}$ is given by the average over all shifts in the reference frame of each spike train
\begin{equation}
	\vec{s^{\lambda}}=\sum_{\alpha} \frac{\vec{s^{\alpha}}}{N}
\end{equation}
and thus as the average of the columns of the spike time difference matrix.

\end{appendix}

%\bibliography{Kreuz_Bibliography}
%\bibliographystyle{elsarticle-harv}
\bibliographystyle{elsarticle-num}

\end{document}